\documentclass[%
 aip,
 amsmath,amssymb,
 reprint,%
]{revtex4-1}

\usepackage{graphicx}% Include figure files
\usepackage{dcolumn}% Align table columns on decimal point
\usepackage{bm}% bold math
\usepackage[utf8]{inputenc}
\usepackage[T1]{fontenc}
\usepackage{mathptmx}
\usepackage{etoolbox}

\newcommand{\pd}[2]{\frac{\partial #1}{\partial #2}}
\newcommand{\ave}[1]{\left\langle #1 \right\rangle}

\makeatletter
\def\@email#1#2{%
 \endgroup
 \patchcmd{\titleblock@produce}
  {\frontmatter@RRAPformat}
  {\frontmatter@RRAPformat{\produce@RRAP{*#1\href{mailto:#2}{#2}}}\frontmatter@RRAPformat}
  {}{}
}%
\makeatother
\begin{document}

\preprint{AIP/123-QED}

\title[Spectral analysis on turbulent momentum and heat transfers in PCT]{Spectral analysis on dissimilarity between turbulent momentum and heat transfers in plane Couette turbulence}
% Force line breaks with \\
\author{T. Kawata}
 \affiliation{ 
Keio University, Hiyoshi 3-14-1, Yokohama, 223-8522 Kanagawa, Japan
}%  \altaffiliation[Also at ]{Physics Department, XYZ University.}%Lines break automatically or can be forced with \\
 \email{kawata@keio.jp}
 
\author{T. Tsukahara}%
 \affiliation{ 
Tokyo University of Science, Yamazaki 2641, Noda, 278-8510 Chiba, Japan
%\\This line break forced with \textbackslash\textbackslash
}%
%  \email{tsuka@rs.tus.ac.jp}

\date{\today}% It is always \today, today,

\begin{abstract}
Nonlinear interaction between different scales in turbulence results in both interscale and spatial transport of turbulent energy, and the role of such scale interactions in turbulent heat transfer mechanism is also of practical importance from the engineering viewpoint. In the present study, we investigate a turbulent plane Couette flow with passive-scalar heat transfer at the Prandtl number 0.71, in order to address the similarity/difference between the scale interactions in the velocity and temperature fields. The constant-temperature-difference boundary condition is used so that the mean velocity and temperature profiles are similar, and then the roles of interscale and spatial transports are compared for the spectral transport budgets of the turbulent energies and temperature-related statistics. We show that turbulent heat transfer occurs at relatively small streamwise length scales compared to momentum transfer, although molecular diffusion is more significant in the temperature field as the Prandtl number is less than 1. Detailed analysis on the transport budgets of the temperature-related spectra shows that scale interactions in temperature field supply more energy to small scales than those in velocity field. Such significant temperature cascade leads to more energetic temperature fluctuation at small scales, which eventually results in the dissimilarity between the spectra of turbulent heat and momentum transfers.
\end{abstract}

\maketitle

% \begin{quotation}
% The ``lead paragraph'' is encapsulated with the \LaTeX\ 
% \verb+quotation+ environment and is formatted as a single paragraph before the first section heading. 
% (The \verb+quotation+ environment reverts to its usual meaning after the first sectioning command.) 
% Note that numbered references are allowed in the lead paragraph.
% %
% The lead paragraph will only be found in an article being prepared for the journal \textit{Chaos}.
% \end{quotation}

\section{\label{sec:intro} Introduction}
% First-level heading:\protect\\ The line break was forced \lowercase{via} \textbackslash\textbackslash

Nonlinear interaction between different scales is the important aspect of turbulence, by which turbulent energy is transported across different scales and spatial diffusions of momentum and heat are enhanced. In wall turbulence at certainly high Reynolds numbers, the flow involves relatively small-scale coherent structures in the near-wall region and the large-scale structures located in the logarithmic region or further out (see, for example, Ref.~\onlinecite{smits_2011} and references therein), and nonlinear interaction between such inner and outer structures has also been pointed out in a number of studied in the last decades~\cite{mathis_2009a,bernardini_2011,pirozzoli_2011,dogan_2019}. 

Motivated by such observations of coherent structures and their interactions, there has been intensive efforts in recent years to investigate the scale interactions in wall turbulence by quantifying turbulent energy transfers between different scales~\cite{cimarelli_2016,mizuno_2016,mamatsashvili_2016,cho_2018,kawata_2018,bauer_2019,hamba_2019,lee_2019,motoori_2019,motoori_2020,chin_2021,gatti_2021,hwang_2021,kawata_2021}. Interestingly, both forward (from larger to smaller scales) and inverse (from smaller to larger scales) turbulent energy cascades have been observed in these studies, and the relations of such interscale energy transfers to the dynamics of coherent structures, such as the self-sustaining cycles of each inner and outer structure\cite{cho_2018,hamba_2019,wang_2021,kawata_2021}, the inner-outer interactions\cite{cho_2018,kawata_2018,kawata_2021,hwang_2021,gatti_2021,mateling2022,chan2022,zhou2022}, merging/splitting of wall-attached hair-pin vortices\cite{chin_2021}, etc.~have been discussed. 

Although only the interscale energy transfer was focused on in the earlier studies mentioned above, the effect of nonlinear scale interactions to enhance spatial diffusions is also an important feature of turbulence, particularly in terms of engineering point of view. One of the most desired controls is the suppressing momentum transfer with enhancing heat transfer at the same time, but such dissimilar control of turbulence is usually difficult as the heat transfer is basically passive to the turbulent fluid motions and therefore there is a certain analogy between the turbulent heat and momentum transfers. 

Some earlier studies, however, demonstrated the possibility of the dissimilar control of turbulent heat and momentum transfers. Hasegawa \& Kasagi\cite{hasegawa_2011} and Yamamoto {\it et al}.\cite{yamamoto_2013} introduced suboptimal control theory and found that the dissimilar control works even in flows where the averaged energy and momentum transport equations have an identical form, and the optimal input mode exhibits a streamwise traveling-wave-like property. Successful dissimilar controls by traveling-wave-like input from the wall have been actually reported by some numerical studies using blowing/suction from the wall\citep{higashi_2011,kaithakkal_2020,mamori_2021} or wall deformation\cite{uchino_2017}. The essential mechanism of such dissimilar transfer control is the different sensitivities of velocity and temperature fields to the control input from the wall, which results from the fact that velocity is a divergence-free vector quantity whereas temperature is a scalar\cite{hasegawa_2011}. Deepening our understanding of such differences in the behaviours of velocity and temperature fields may allow us to further develop the dissimilar flow control schemes.

Despite numerous studies on wall turbulence with passive-scalar temperature field reported in the past including aforementioned attempts for the dissimilar flow control, less attention has been paid to the spectra and cascades of fluctuating temperature (or passive scalar). Among such few studies, Antonia \& Abe\cite{antonia_2009} compares the spectra of the turbulent kinetic energy and the scalar temperature fluctuation in turbulent channel flows and reported no significant difference between those energy spectra, but their dissipation spectra display qualitatively different behaviours. Hane~{\it et al}.\cite{hane_2006} simulated a turbulent plane Couette flow with heat transfer and showed that the spectra of the streamwise turbulent energy and temperature fluctuation indicate their peaks at different streamwise wavelengths. These studies suggest that the scales dominating the heat and momentum transfers may differ in spite of the inherent analogy.  As the turbulent heat and momentum transfers are the results of fluid motions in a wide range of scale, investigating their mechanisms scale-by-scale should help us, for example, develop new control strategies for heat transfer enhancement. 

In the present study, we investigate the transport equation budgets of the temperature-related spectra based on the direct numerical simulation data of wall turbulence with passive-scalar temperature field, with a particular focus on the roles of the interscale and spatial transports at each scale. Turbulent plane Couette flow is chosen for the test case, because very-large-scale structures filling up the entire channel appear even at relatively low Reynolds numbers\cite{bech_1995,komminaho_1996,kitoh_2005,tsukahara_2006,kitoh_2008,avsarkisov_2014,orlandi_2015,lee_2018} and therefore the scale separation between the inner and outer structures is relatively clear as compared to other canonical wall-bounded flow configurations at similar Reynolds numbers. The constant-temperature-difference boundary condition is applied to the temperature field so that the mean velocity and temperature profiles are similar to each other, and then fluctuating velocity and temperature fields are compared focusing on similarity/difference in the interscale and spatial energy transports caused by scale interactions. 

\section{\label{sec:setup} Numerical simulation and data analysis}

\subsection{Direct Numerical Simulation of Plane Couette Flow with Passive-Scalar Temperature Transport}

In the configuration of the plane Couette flow simulated in this study, a flow between two parallel flat plates separated by distance~$h^\ast$ (hereafter the superscript $^\ast$ indicates dimensional quantities) is driven by translation of the top wall with a constant velocity~$U^\ast_\mathrm{w}$ (the bottom wall is stationary). The $x^\ast$-, $y^\ast$- and $z^\ast$-axes are taken in the streamwise, wall-normal and spanwise directions, and the origin of the coordinates is placed on the stationary bottom wall; the bottom and top walls are located at $y^\ast=0$ and $y^\ast=h^\ast$, respectively. The temperatures of the flat plates are uniform and constant in time with the temperature difference $\Delta T^\ast=T^\ast_\mathrm{t}-T^\ast_\mathrm{b}$ ($T^\ast_\mathrm{t}$ and $T^\ast_\mathrm{b}$ are the temperature of the top and the bottom wall, respectively).  

The governing equations for the fluid flow simulation are the non-dimensional continuity and Navier-Stocks equation for incomplessible fluid flow scaled by $U_\mathrm{w}^\ast$ and $h^\ast$: 
\begin{subequations} \label{eq:NS}
\begin{align}
\nabla \cdot \tilde{\bf u} &= {\bf 0}, \\
\pd{\tilde{\bf u}}{t} + (\tilde{\bf u}\cdot \nabla) \tilde{\bf u}&= -\nabla \tilde{p} + \frac{1}{Re_\mathrm{w}} {\nabla}^2 \tilde{\bf u}.
\end{align}
\label{eq:fluid}
\end{subequations}
Here the tilde ($\tilde{\;\;}$) indicates an instantaneous quantity including both the mean and the fluctuating components, and $\tilde{\bf u}$ and $\tilde{p}$ are the instantaneous velocity vector and pressure of the fluid, respectively. $Re_\mathrm{w}=U_\mathrm{w}^\ast h^\ast/\nu^\ast$ is the Reynolds number ($\nu^\ast$ is the kinematic viscosity).  
The governing equation of the fluctuating temperature field is
\begin{align}
\pd{\tilde{\theta}}{t} + (\tilde{\bf u}\cdot \nabla) \tilde{\theta} = \frac{1}{Re_\mathrm{w} Pr} {\nabla}^2 \tilde{\theta}, \label{eq:theta}
\end{align}
where $Pr$ is the Prandtl number and $\tilde{\theta}$ is the instantaneous non-dimensional fluid temperature defined as
\begin{align}
\tilde{\theta} = \frac{\tilde{\theta}^\ast}{\Delta T^\ast} \quad
(\tilde{\theta}^\ast = \tilde{T}^\ast-T_b^\ast).
\end{align}
Here $\tilde{T}^\ast$ is the dimensional instantaneous fluid temperature. In the present computation, the values of the Reynolds and Prandtl numbers are $Re_\mathrm{w}=8600$ and $Pr=0.71$, respectively.

The present computation is based on our previous study on turbulent plane Couette flow\cite{kawata_2021}. Hence, the numerical procedures to solve the governing equations for fluid motion~(\ref{eq:NS}) are the same, and the equation for the temperature field~(\ref{eq:theta}) is additionally solved with the same numerically schemes as used for Eqs.~(\ref{eq:fluid}). The details of the numerical procedures have already been described elsewhere~(for example, Ref.~\onlinecite{tsukahara_2006}), and are therefore only given here briefly. The governing equations~(\ref{eq:fluid}) and (\ref{eq:theta}) are solved in physical space with spatial discretization of the finite difference method, while the pressure Poisson equation is solved in Fourier space. Central difference schemes with fourth and second order are used for the wall-parallel ($x$- and $z$-) and the wall-normal ($y$-) directions, respectively. Periodic boundary condition is applied to both the velocity and temperature fields for $x$- and $z$-directions, and similar constant-difference conditions are applied to the velocity and temperature fields on the walls:
\begin{subequations}
\begin{align}
    \tilde{\bf u}={\bf 0}, \quad
    \tilde{\theta}=0 \quad \text{at} \quad y=0, \\
    \tilde{\bf u}=(1,0,0), \quad
    \tilde{\theta}=1 \quad \text{at} \quad  y=1.
\end{align}
\end{subequations}

Due to the existence of the very-large-scale structures in the turbulent plane Couette flows one needs to use very large computational domains to exclude the confinement effects on the simulated flow field by the periodic boundary conditions\citep{komminaho_1996,tsukahara_2006}. In our previous study~\citep{kawata_2021} we carefully examined the domain-size effect, and the computational conditions, such as the domain size and the spatial resolutions in $x$-, $y$- and $z$-directions, in the present numerical simulation are the same as in the Case~A1 in the previous study (see Table~\ref{tab:num}), which was confirmed to give computational results in a quite good agreement with those obtained with an extremely large domain ($L_x^\ast=96h$ and $L_z^\ast=12.8h$). With the above-described numerical setup the outer-scale Reynolds number $Re_\mathrm{w}=8600$ results in the friction Reynolds number $Re_\tau=u_\tau^\ast \delta^\ast/\nu^\ast=126.0$ (here $\delta^\ast=h^\ast/2$ is the channel half height and $u^\ast_\tau$ is the friction velocity defined as $u^\ast_\tau=\sqrt{\tau^\ast_\mathrm{w}/\rho^\ast}$ with $\tau^\ast_\mathrm{w}$ and $\rho^\ast$ being the wall shear stress and the fluid density, respectively), as also given in Table~\ref{tab:num}.

\begin{table}
    \caption{Computational conditions: the streamwise and spanwise domain lengths, the number of grid points, the spatial resolutions, and the friction Reynolds number $Re_\tau$. The outer Reynolds number and the Prandtl number are $Re_\mathrm{w}=8600$ and $Pr=0.71$, respectively.}
    \label{tab:num}
    \begin{ruledtabular}
    \begin{tabular}{ccccc}
        $(L_x^\ast,L_z^\ast)$ & $(N_x,N_y,N_z)$ & $(\Delta x^+,\Delta z^+)$ & $(\Delta y^+_\mathrm{max},\Delta y^+_\mathrm{min})$ & $Re_\tau$\\
        $(24h,12.8h)$ & $(512,96,512)$ & $(11.8,6.31)$ & $(6.16,0.268)$ & $126.1$
    \end{tabular}
    \end{ruledtabular}
\end{table}

\subsection{Spectral Analysis on the Reynolds-Stress Transport and Turbulent Heat Transfers}

In the present study, the spectral transports of the temperature fluctuation $\ave{\theta^2}$ and the turbulent heat fluxes $\ave{u_i \theta}$ are analysed. The transport equations of the temperature-related spectra are derived similarly to those of the Reynolds stress spectra. In this section, the derivation of the transport equations of the Reynolds stress spectra is briefly reviewed and that for the temperature-related turbulence statistics are defined. In the following, the instantaneous velocities in $x$-, $y$- and $z$-directions are denoted as $\tilde{u}=U+u$, $\tilde{v}=v$, and $\tilde{w}=w$, and the instantaneous temperature is also decomposed as $\tilde{\theta}=\varTheta+\theta$. The uppercase letters $U=\ave{\tilde{u}}$ and $\varTheta=\ave{\tilde{\theta}}$ are the mean streamwise velocity and temperature (here, $\ave{}$ stands for averaging in $x$- and $z$-directions and in time), respectively, and the lowercase letters without the tilde represent the fluctuations around the mean values. Note here $\ave{\tilde{v}}=\ave{\tilde{w}}=0$.

\subsubsection{Transport of Reynolds stresses}

The transport equation of the Reynolds stress $\ave{u_i u_j}$ is expressed as
\begin{subequations} \label{eq:rste}
\begin{equation}
\left(\pd{}{t}+U_k\pd{}{x_k}\right) \ave{u_i u_j} = P_{ij}-\varepsilon_{ij}+\Phi_{ij}+D^\nu_{ij}+D^t_{ij}, 
\end{equation}
where the terms on the right-hand side are the production ($P_{ij}$), the viscous dissipation ($\varepsilon_{ij}$), the pressure work ($\Phi_{ij}$), the viscous diffusion ($D^\nu_{ij}$), and the turbulent diffusion ($D^t_{ij}$) terms, which are defined as
\begin{align}
P_{ij}&=-\ave{u_i u_k}\pd{U_j}{x_k}-\ave{u_j u_k}\pd{U_i}{x_k}, \\
\varepsilon_{ij} &= \frac{2}{Re_\mathrm{w}} \ave{\pd{u_i}{x_k}\pd{u_j}{x_k}},\\
\Phi_{ij} &= -\ave{\pd{p}{x_i}u_j}-\ave{\pd{p}{x_j}u_i}, \\
D^\nu_{ij} &= \frac{1}{Re_\mathrm{w}} \pd{^2 \ave{u_i u_j}}{x_k^2}, \\
D^t_{ij} &= -\pd{\ave{u_i u_j u_k}}{x_k}.
\end{align}
\end{subequations}
The pressure-work term $\Phi_{ij}$ may be decomposed into the pressure-strain correlation $\Pi_{ij}$ and the pressure diffusion $D^p_{ij}$:
\begin{subequations} \label{eq:Pi_ij}
\begin{align}
    \Phi_{ij} &= \Pi_{ij} + D^p_{ij}, \\
    \mathrm{where} \quad \Pi_{ij} &= \ave{p \left( \pd{u_i}{x_j} + \pd{u_j}{x_i}\right)}, \\
                   D^p_{ij} &= - \pd{}{x_k} \left( \ave{u_j p} \delta_{ik} + \ave{u_i p} \delta_{jk}\right).
\end{align} 
\end{subequations}
Here, $\delta_{ij}$ is the Kronecker's delta.

Now, we consider a decomposition of the fluctuating components into their large- and small-scale parts 
\begin{align} \label{eq:dec}
    {\bf u}={\bf u}^L + {\bf u}^S, \quad \theta=\theta^L+\theta^S,
\end{align}
where the cross correlations between the large- and small-scale part of any quantities are zero:
\begin{align}
    \ave{u_i^L u_j^S} = \ave{u_j^L u_i^S} = 0, \nonumber \\
    \ave{\theta^L \theta^S} = \ave{u_i^L \theta^S} = \ave{u_i^S \theta^L} = 0. \nonumber
\end{align}
Such a decomposition is possible based on orthogonal mode decompositions such as the Fourier mode decomposition and the proper-orthogonal-mode decomposition. We adopt a decomposition based on the streamwise or spanwise Fourier modes at a certain cutoff wavenumber $k_c$. By the decomposition the Reynolds stress $\ave{u_i u_j}$ is decomposed simply into their large- and small-scale parts as
\begin{align} \label{eq:rsdec}
    \ave{u_i u_j} = \ave{u_i^L u_j^L} + \ave{u_i^S u_j^S}.
\end{align}
With a similar manner to the ``full'' Reynolds stress equation~(\ref{eq:rste}), one can obtain the transport equations of the large- and small-scale parts of $\ave{u_i u_j}$ (the details of the derivation are found in Refs.~\onlinecite{kawata_2019,kawata_2021}) as
\begin{subequations} \label{eq:rstels}
\begin{align}
\frac{{\rm d}\ave{u_i^L u_j^L}}{{\rm d}t} = P^L_{ij}-\varepsilon^L_{ij}+\Phi^L_{ij}+D^{\nu,L}_{ij}+D^{t,L}_{ij} -Tr_{ij}, \\
\frac{{\rm d}\ave{u_i^S u_j^S}}{{\rm d}t} = P^S_{ij}-\varepsilon^S_{ij}+\Phi^S_{ij}+D^{\nu,S}_{ij}+D^{t,S}_{ij}+Tr_{ij}. 
\end{align}
\end{subequations}
Here, ${\rm d}/{\rm d}t=\partial /\partial t + U_k \partial/\partial x_k$, and the terms on the right-hand sides of Eqs.~(\ref{eq:rstels}a, b) are the large- and small-scale parts of their counterparts in Eq.~(\ref{eq:rste}), except for $Tr_{ij}$. In particular, the first four terms in Eq.~(\ref{eq:rste}) (i.e., the production, the viscous dissipation, the pressure work, and the viscous diffusion) are simply decomposed into their large- and small-scale parts similarly in Eq.~(\ref{eq:rsdec}): those in Eq.~(\ref{eq:rstels}a) are defined as in Eq.~(\ref{eq:rste}b--e) with $u_i$ and $u_j$ replaced by $u_i^L$ and $u_j^L$ (for those in Eq.~(\ref{eq:rstels}b), replaced by $u_i^S$ and $u_j^S$). 

On the other hand, the turbulent diffusion is decomposed as  
\begin{align}
D^{t,L}=-\pd{\ave{u_i u_j u_k}^L}{x_k}, \quad 
D^{t,S}=-\pd{\ave{u_i u_j u_k}^S}{x_k},
\end{align}
where $\ave{u_i u_j u_k}^L$ and $\ave{u_i u_j u_k}^S$ are the large- and small-scale part of the triple velocity correlation $\ave{u_i u_j u_k}$ defined as
\begin{subequations} \label{eq:dectri}
\begin{align}
\ave{u_i u_j u_k}^L=\ave{u_i^L u_j^L u_k}+\ave{u_i^S u_j^L u_k^S}+\ave{u_i^L u_j^S u_k^S},\\
\ave{u_i u_j u_k}^S=\ave{u_i^S u_j^S u_k}+\ave{u_i^L u_j^S u_k^L}+\ave{u_i^S u_j^L u_k^L}.
\end{align}
\end{subequations}
The last term in Eqs.~(\ref{eq:rstels}a, b), $Tr_{ij}$, is defined as 
\begin{align} \label{eq:Trij}
Tr_{ij} = 
-\ave{u_i^S u_k^S \pd{u_j^L}{x_k}}-\ave{u_j^S u_k^S \pd{u_i^L}{x_k}} \nonumber \\
+\ave{u_i^L u_k^L \pd{u_j^S}{x_k}}+\ave{u_j^L u_k^L \pd{u_i^S}{x_k}},
\end{align}
and as can be seen from Eqs.(\ref{eq:rstels}a, b) this term appears in both equations with opposite signs, which means that $Tr_{ij}$ represents the energy exchange between the large- and small-scale velocity fields across the cutoff wavenumber $k_c$. 

It is worth pointing out here that the turbulent transport terms ($D^{t,L}_{ij}$ and $D^{t,S}_{ij}$) and the interscale energy flux term ($Tr_{ij}$) includes both the large- and small-scale part of fluctuating velocities, which is in contrast to the other terms in the the transport equations consisting of only either ${\bf u}^L$ or ${\bf u}^S$. This is because both the turbulent transport and the interscale energy flux terms are the third-order moments of fluctuating velocities (and the velocity gradients), while the other terms are of the second moments. As $D^{t,L}_{ij}$, $D^{t,S}_{ij}$, and $Tr_{ij}$ include both large- and small-scale parts of the velocity field, these terms can be interpreted to represent the energy-transport effects by interactions between different scales: the turbulent transport ($D^{t,L}_{ij}$ and $D^{t,S}_{ij}$) indicates the energy transport in physical space caused by scale interactions, while the interscale energy flux ($Tr_{ij}$) is the energy transport between different scales. 

As the large-scale part of the Reynolds stresses $\ave{u_i^L u_j^L}$ and the Reynolds stress spectra $E_{ij}$ are related as 
\begin{align} \label{eq:eij_uiuj}
    E_{ij} = \pd{\ave{u_i^L u^L_j}}{k_c} \left(=-\pd{\ave{u_i^S u_j^S}}{k_c} \right),
\end{align}
the transport equations of the Reynolds-stress spectra can, therefore, be derived by differentiating both sides of Eq.~(\ref{eq:rstels}a) with respect to $k_c$ as
\begin{align} \label{eq:eijte}
\frac{{\rm d} E_{ij}}{{\rm d}t} =
pr_{ij}-\xi_{ij}+\phi_{ij}+d^\nu_{ij}+d^t_{ij}+tr_{ij},
\end{align}
where the terms on the right-hand side are the $k_c$-derivatives of the counterparts in Eq.~(\ref{eq:rstels}a). The first five terms represent the spectra of the corresponding terms of Eq.(~\ref{eq:rste}a): the spectra of the production ($pr_{ij}$), the viscous dissipation ($\xi_{ij}$), the pressure-work ($\phi_{ij}$), the viscous diffusion $d^\nu_{ij}$, and the turbulent transport $d^t_{ij}$. The interscale transport term~$tr_{ij}$ represents the energy gain/loss by the interscale flux $Tr_{ij}$ at each wavenumebr. In particular, the spectra of the production ($pr_{ij}$) and the viscous diffusion ($d^\nu_{ij}$) are expressed based on the Reynolds stress spectra as
\begin{subequations}
\begin{align}
    pr_{ij} &= -E_{ik} \pd{U_j}{x_k} - E_{jk}\pd{U_i}{x_k}, \\
    d^\nu_{ij} &= \frac{1}{Re_\mathrm{w}} \pd{^2 E_{ij}}{x_k^2}.
\end{align}
\end{subequations}
% The spectra of viscous dissipation ($\xi_{ij}$) and the pressure work ($\phi_{ij}$) are defined as the cospectra between different velocity gradients and the pressure-gradient-velocity cospectra, respectively, and 
According to Eq.~(\ref{eq:Pi_ij}), the pressure-work spectrum $\phi_{ij}$ may be decomposed into the spectra of $\Pi_{ij}$ and $D^p_{ij}$ as 
\begin{align} \label{eq:pi_ij}
    \phi_{ij} = \pi_{ij} + d^p_{ij}.
\end{align}
Here, $\pi_{ij}$ is the cospectrum between the pressure and the strain rate, representing the inter-component energy transfers between different components of velocity fluctuation at each scale, and $d^p_{ij}$ consists of the velocity-pressure cospectra, representing spatial energy transport by pressure fluctuation at each scale. 

The turbulent spatial transport $d^t_{ij}$ and the interscale transport $tr_{ij}$, which are of particular interest in the present study, are respectively defined as
\begin{subequations} \label{eq:eijk}
\begin{align}
d^t_{ij}=-\pd{E_{ijk}}{x_k}, \quad\quad tr_{ij} = - \pd{Tr_{ij}}{k_c}.
\end{align}
Here, $E_{ijk}$ is the spectra of triple velocity correlations $\ave{u_i u_j u_k}$ defined as
\begin{align}
E_{ijk} = \pd{\ave{u_i u_j u_k}^L}{k_c} \left(= -\pd{\ave{u_i u_j u_k}^S}{k_c} \right),
\end{align}
\end{subequations}
similarly to the definition of the Reynolds stress spectra, Eq.~(\ref{eq:eij_uiuj}). As the physical meaning of $\ave{u_i u_j u_k}$ is the spatial transport flux of the Reynolds stress $\ave{u_i u_j}$ in the $x_k$-direction caused by the velocity fluctuation $u_k$, their spectra $E_{ijk}$ represent the spatial flux of the Reynolds stress $\ave{u_i u_j}$ in the $x_k$-direction at each wavenumber. Again, note that the spatial transport represented by $E_{ijk}$ is caused by interactions between different scales as indicated by Eqs.~(\ref{eq:dectri}). 

\subsubsection{Transport of temperature fluctuation and velocity-temperature correlations}

Now we derive the spectral transport equations of the temperature-related turbulence statistics. The transport equation of the temperature fluctuation $\ave{\theta^2}$ is written similarly to the Reynolds stress equation (\ref{eq:rste}) as
\begin{subequations}\label{eq:ttte}
\begin{equation}
\left(\pd{}{t}+U_k\pd{}{x_k}\right) \ave{\theta^2} = P_{\theta \theta}-\varepsilon_{\theta \theta}+D^\nu_{\theta \theta}+D^t_{\theta \theta} ,
\end{equation}
with the production ($P_{\theta\theta}$), the viscous dissipation ($\varepsilon_{\theta\theta}$), the viscous diffusion ($D^\nu_{\theta\theta}$), and the turbulent transport ($D^t_{\theta\theta}$) terms defined as
\begin{align}
P_{\theta \theta} &= - 2\ave{\theta u_k}\pd{\varTheta}{x_k}, \\
\varepsilon_{\theta\theta} &= 
\frac{2}{Re_\mathrm{w}Pr} \ave{\pd{\theta}{x_k} \pd{\theta}{x_k}}, \\
D^\nu_{ij} &= \frac{1}{Re_\mathrm{w}Pr} \pd{^2 \ave{\theta^2}}{x_k^2}, \\
D^t_{\theta\theta} &= -\pd{\ave{\theta^2 u_k}}{x_k}.
\end{align}
\end{subequations}
It is noteworthy here that any term corresponding to $\Phi_{ij}$ in Eq.~(\ref{eq:rste}a) does not exist in Eq.~(\ref{eq:ttte}a), since the energy equation (\ref{eq:theta}) does not have pressure-related term. 

The transport equations of the turbulent heat fluxes $\ave{u_i \theta}$ are also similarly obtained as
\begin{subequations}\label{eq:tvte}
\begin{equation}
\left(\pd{}{t}+U_k\pd{}{x_k}\right) \ave{u_i \theta} = P_{i \theta}-\varepsilon_{i \theta}+\Phi_{i\theta}+D^\nu_{i \theta}+D^t_{i \theta},
\end{equation}
with the production ($P_{i \theta}$), the viscous dissipation ($\varepsilon_{i \theta}$), the pressure-gradient-temperature correlations ($\Phi_{i \theta}$), the viscous diffusion ($D^\nu_{i \theta}$) and the turbulent transport ($D^t_{i \theta}$) terms:
\begin{align}
P_{i\theta}&=-\ave{u_k \theta} \pd{U_i}{x_k} - \ave{u_i u_k}\pd{\varTheta}{x_k}, \\
\varepsilon_{i\theta} &= \frac{1}{Re_\mathrm{w}} \left( 1 + \frac{1}{Pr} \right) \ave{\pd{u_i}{x_k} \pd{\theta}{x_k}}, \\
\Phi_{i\theta} &= -\ave{\pd{p}{x_i} \theta}, \\
D^\nu_{i\theta} &= \frac{1}{Re_\mathrm{w}} \pd{}{x_k}
\left( \ave{\pd{u_i}{x_k}\theta} + \frac{1}{Pr}\ave{\pd{\theta}{x_k}u_i}
\right), \\
D^t_{i\theta} &= -\pd{\ave{u_i\theta u_k}}{x_k}.
\end{align}
\end{subequations}

Based on the decomposition given by Eq.~(\ref{eq:dec}), the temperature fluctuation and the turbulent heat fluxes are also split into their large- and small-scale parts, similarly to the Reynolds stress decomposition (\ref{eq:rsdec}), as
\begin{subequations}
\begin{align}
    \ave{\theta^2} = \ave{\theta^L \theta^L}+\ave{\theta^S \theta^S},\\ 
    \ave{u_i \theta} = \ave{u_i^L \theta^L} + \ave{u_i^S \theta^S},
\end{align}
\end{subequations}
and therefore the transport equations of their large- and small-scale parts can be derived similarly to Eqs.~(\ref{eq:rstels}). Then, these large- and small-scale parts of the temperature-related statistics are related to their spectra, similarly to Eq.~(\ref{eq:eij_uiuj}), as
\begin{align}
    E_{\theta \theta} = \pd{\ave{\theta^L \theta^L}}{k_c}, \quad
    E_{i \theta} = \pd{\ave{u_i^L \theta^L}}{k_c},
\end{align}
and hence the transport equations of these temperature-related spectra can also be obtained by differentiating the equations of $\ave{\theta^L \theta^L}$ and $\ave{u_i^L \theta^L}$. The transport equation of the temperature-fluctuation spectrum $E_{\theta \theta}$ is expressed as 
\begin{align} \label{eq:ettte}
\frac{{\rm d} E_{\theta \theta}}{{\rm d}t}  &= pr_{\theta\theta} - \xi_{\theta\theta} + d^\nu_{\theta\theta} + d^t_{\theta\theta} + tr_{\theta\theta}
\end{align}
with the terms on the right-hand side representing the spectra of the counterparts in Eq.~(\ref{eq:ttte}). 
% the production and the viscous diffusion spectra are defined as
% \begin{align}
%     pr_{\theta\theta} &= -2 E_{u_k \theta} \pd{\varTheta}{x_k} \\
%     d^\nu_{\theta\theta} &= \frac{2}{Re_\mathrm{w} Pr} \pd{^2 E_{\theta\theta}}{x_k}, 
% \end{align}
% and the dissipation spectra $\xi_{\theta \theta}$ consist of the cospectra of temperature gradients. 
The interscale and the turbulent spatial transport terms are defined as
\begin{subequations} \label{eq:si_tt}
\begin{align}
tr_{\theta\theta} = - \pd{Tr_{\theta\theta}}{k_c}, \quad\quad d^t_{\theta\theta} = - \pd{E_{\theta\theta k}}{x_k}, 
\end{align}
where $Tr_{\theta \theta}$ and $E_{\theta \theta k}$ are the interscale and the spectral spatial fluxes of the temperature fluctuation defined as
\begin{align} 
Tr_{\theta\theta} &= -2\ave{\theta^S u_k^S \pd{\theta^L}{x_k}} -2\ave{\theta^L u_k^L \pd{\theta^S}{x_k}}, \\ 
E_{\theta\theta k} &= \pd{\ave{\theta^2 u_k}^L}{k_c} = -\pd{\ave{\theta^2 u_k}^S}{k_c},
\end{align}
with $\ave{\theta^2 u_k}^L$ and $\ave{\theta^2 u_k}^S$ being the large- and small-scale parts of the spatial flux of temperature fluctuation $\ave{\theta^2 u_k}$:
\begin{align}
\ave{\theta^2 u_k}^L &= \ave{\theta^L \theta^L u_k} + 2\ave{\theta^L \theta^S u_k^S}, \\
\ave{\theta^2 u_k}^S &= \ave{\theta^S \theta^S u_k} + 2\ave{\theta^L \theta^S u_k^L}.
\end{align}
\end{subequations}
Similarly, the transport equation of the turbulent heat flux spectrum $E_{i\theta}$ is obtained as
\begin{align} \label{eq:eitte}
 \frac{{\rm d} E_{i \theta}}{{\rm d}t} &= pr_{i\theta} - \xi_{i\theta} +\phi_{i\theta}+ d^\nu_{i\theta} + d^t_{i\theta} + tr_{i\theta},
\end{align}
and the terms on the right-hand side (except $tr_{i\theta}$) are, again, the spectra of their counterparts in Eq.~(\ref{eq:tvte}).
% the production and the viscous diffusion spectra are defined as
% \begin{align}
%     pr_{i\theta} = -E_{k\theta} \pd{U_i}{x_k} - E_{ik} \pd{\varTheta}{x_k},
% \end{align}
% the dissipation spectra $\xi_{i\theta}$ is the cospectra between the temperature- and velocity-velocity gradients, the pressure-work term $\phi_{i\theta}$ the pressure-gradient-temperature cospectra, and the viscous diffusion term $d^\nu_{ij}$ is the sum of the velocity-gradient-temperature cospectra and the temperature-gradient-velocity cospectra. 
The interscale and the turbulent spatial transport terms are defined in the same manners as in other spectral transport equations: 
\begin{subequations} \label{eq:si_ut}
\begin{align}
tr_{i\theta} = - \pd{Tr_{i\theta}}{k_c}, \quad\quad d^t_{i\theta} = - \pd{E_{i\theta k}}{x_k},
\end{align}
where the interscale flux $Tr_{i\theta}$ and the spectral spatial flux $E_{i\theta k}$ are respectively defined as 
\begin{align}
Tr_{i\theta} =& -\ave{\theta^S u_k^S \pd{u_i^L}{x_k}} -\ave{u_i^S u_k^S \pd{\theta^L}{x_k}} \nonumber \\
 &+\ave{\theta^L u_k^L \pd{u_i^S}{x_k}} +\ave{u_i^L u_k^L \pd{\theta^S}{x_k}}, \\ 
E_{i\theta k} =& \pd{\ave{u_i \theta u_k}^L}{k_c} = -\pd{\ave{u_i \theta u_k}^S}{k_c}.
\end{align}
Here, $\ave{u_i \theta u_k}^L$ and $\ave{u_i \theta u_k}^S$ are the large- and small-scale parts of $\ave{u_i \theta u_k}$ defined as
\begin{align}
\ave{u_i \theta u_k}^L = \ave{u_i^L \theta^L u_k} + \ave{u_i^L \theta^S u_k^S} + \ave{u_i^S \theta^L u_k^S}, \\
\ave{u_i \theta u_k}^S = \ave{u_i^S \theta^S u_k} + \ave{u_i^S \theta^L u_k^L} + \ave{u_i^L \theta^S u_k^L}.
\end{align}
\end{subequations}

\section{Results}

\subsection{Basic Statistics and Spectra of Turbulent Heat and Momentum Transfers}

\begin{figure}
    \centering
    \includegraphics[width=0.85\hsize]{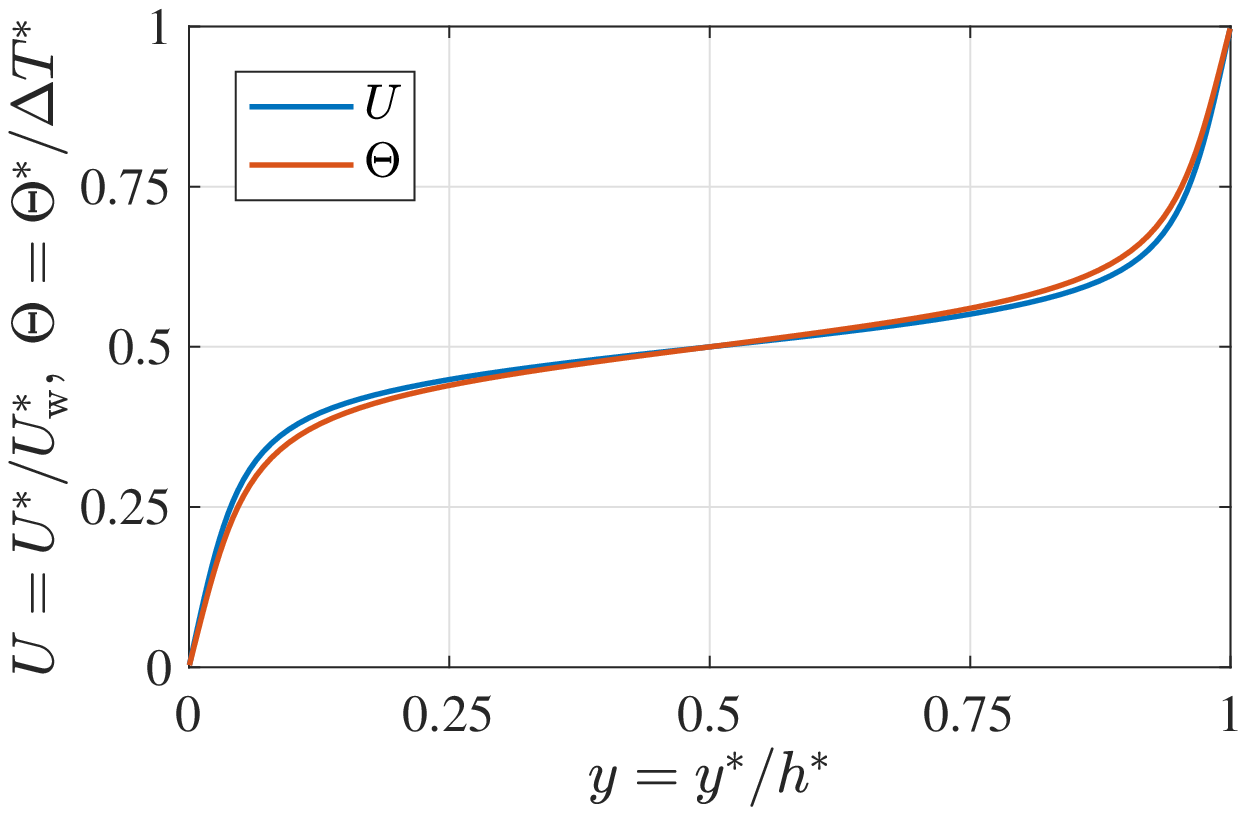}
    \includegraphics[width=0.85\hsize]{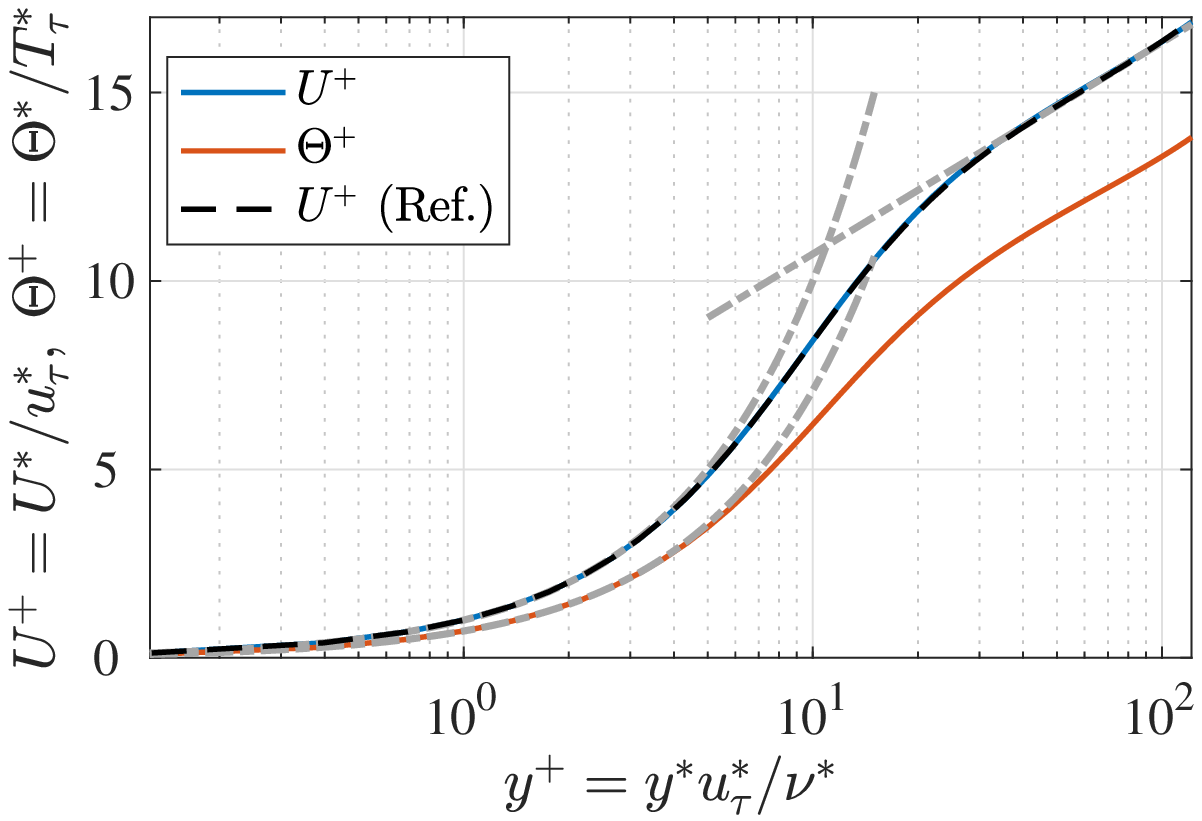}
    \caption{Profiles of mean streamwise velocity $U$ and mean temperature $\varTheta$ in (top) outer and (bottom) inner scaling. In the bottom panel, the black dashed line presents reference data of mean streamwise velocity obtained in Ref.~\onlinecite{kawata_2021} with an extremely large computational domain $(L_x^\ast,L_z^\ast)=(96h, 12.8h)$, and the grey chained lines indicate $U^+=y^+$ and $U^+=2.44 \ln y^+ +5.1$ for mean velocity profile and $\varTheta^+=Pr y^+$ for mean temperature profile.}
    \label{fig:meanUT}
\end{figure}

Figure \ref{fig:meanUT} presents the profiles of the mean streamwise velocity $U$ and the mean temperature $\varTheta$ in the outer and inner scalings. Here, $T^\ast_\tau$ used for the inner scaling of temperature is the friction temperature defined as 
\begin{align}
T_\tau^\ast=\frac{Q_\mathrm{w}^\ast}{\rho^\ast c_p^\ast u_\tau^\ast}    
\end{align}
where $Q_\mathrm{w}^\ast$ and $c_p^\ast$ are the mean heat flux on the wall and the specific heat at constant pressure, respectively. As shown in the top panel, the profiles of $U$ and $\varTheta$ are quite similar to each other, but the temperature gradient in the vicinity of the wall is certainly smaller than the velocity gradient, which is displayed more clearly by the plot in the inner scaling in the bottom panel. This is attributable to the Prandtl number being smaller than $1$, which means that the molecular diffusion effect is more significant in the temperature field. In the bottom panel the profile of $U^+$ obtained in our previous study\cite{kawata_2021} with an extremely large computational domain ($L_x^\ast=96.0h$, $L_z^\ast=12.8h$) is also given by the black dashed line for reference. As can be seen here, the profile obtained in the present computation is in a good agreement with the reference data. 

\begin{figure}
    \centering
    \includegraphics[width=0.9\hsize]{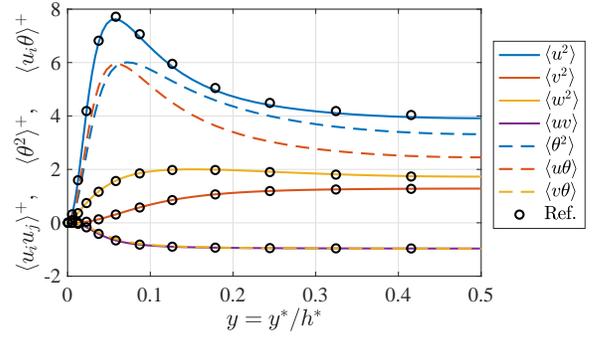}
    \caption{Profiles of (solid lines) the Reynolds stresses $\ave{u_i u_j}$ and (dashed lines) temperature-related turbulent statistics $\ave{\theta^2}$ and $\ave{u_i \theta}$. The values are scaled based on the inner units, $u_\tau^\ast$ and/or $T_\tau^\ast$, and only $0 \ll y \ll 0.5$ is shown. The black circles indicate the profiles of the Reynolds stresses obtained with an extremely large computational domain $(L_x^\ast,L_z^\ast)=(96h,12.8h)$\cite{kawata_2021}. }
    \label{fig:rs}
\end{figure}

Figure~\ref{fig:rs} presents the profiles of the Reynolds stresses $\ave{u^2}$, $\ave{v^2}$, $\ave{w^2}$, and $\ave{uv}$, and the temperature-related statistics: the temperature fluctuation $\ave{\theta^2}$ and the velocity-temperature correlations, i.e., the streamwise and wall-normal turbulent heat fluxes $\ave{u\theta}$ and $\ave{v\theta}$. Note that in the figure only the lower half of the channel $0 \ll y \ll 0.5$ is shown as the statistics are averaged between the upper and lower halves of the channel. Also in this figure the reference data from our previous work~\cite{kawata_2021} are shown (black circles) and the consistency between the present and reference data can be confirmed, indicating that no effect of the finite domain size in the present computation. It is also shown in Fig.~\ref{fig:rs} that the temperature fluctuation $\ave{\theta^2}$ and the streamwise-velocity-temperature correlation $\ave{u\theta}$ give similar profiles to the streamwise velocity fluctuation $\ave{u^2}$, which can be attributed to the similar profiles of the mean streamwise velocity $U$ and the mean temperature $\varTheta$, as shown in Fig.~\ref{fig:meanUT}. The smaller peak magnitudes of the $\ave{\theta^2}$ profile may correspond to the fact that the mean temperature gradient is certainly smaller than the mean velocity gradient in the near-wall region. It is also observed that the peak magnitude of the $\ave{u\theta}$ is as large as those of $\ave{u^2}$ and $\ave{\theta^2}$, which indicates the strong cross correlation between $u$ and $\theta$. 

The correlation $\ave{v\theta}$ is a particularly important temperature-related quantity, as it represents the wall-normal heat transfer by turbulent fluid motions, similarly to the Reynolds shear stress $\ave{uv}$ representing the momentum transfer by turbulence. As shown in Fig.~\ref{fig:rs}, the profiles of $\ave{v\theta}$ and $\ave{uv}$ are almost on top on each other, which indicates a strong similarity between the turbulent momentum and the heat transfers.

\begin{figure*}
    \includegraphics[width=0.8\hsize]{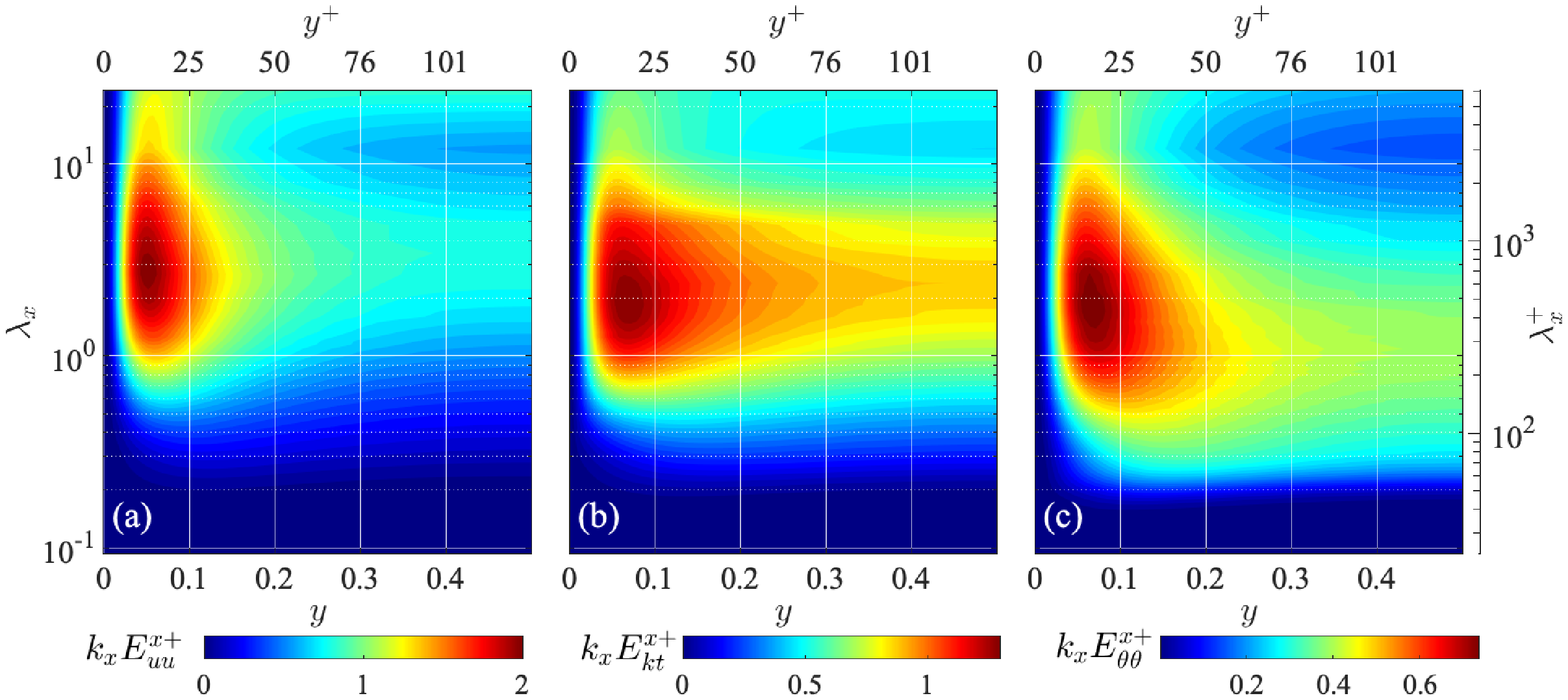}
    \caption{Space-wavelength ($y$-$\lambda_x$) diagrams of premultiplied streamwise one-dimensional spectra of (a) streamwise turbulent energy $k_x E^x_{uu}$, turbulent kinetic energy $k_x E^x_{kt}=k_x(E^x_{uu}+E^x_{vv}+E^x_{ww})/2$, and (c) temperature fluctuation $k_x E^x_
    {\theta \theta}$. The values are scaled by (a,b) $u_\tau^{\ast 2}$ and (c) $T_\tau^{\ast 2}$. }
    \label{fig:ex}
    \vspace{1em}
    \includegraphics[width=0.8\hsize]{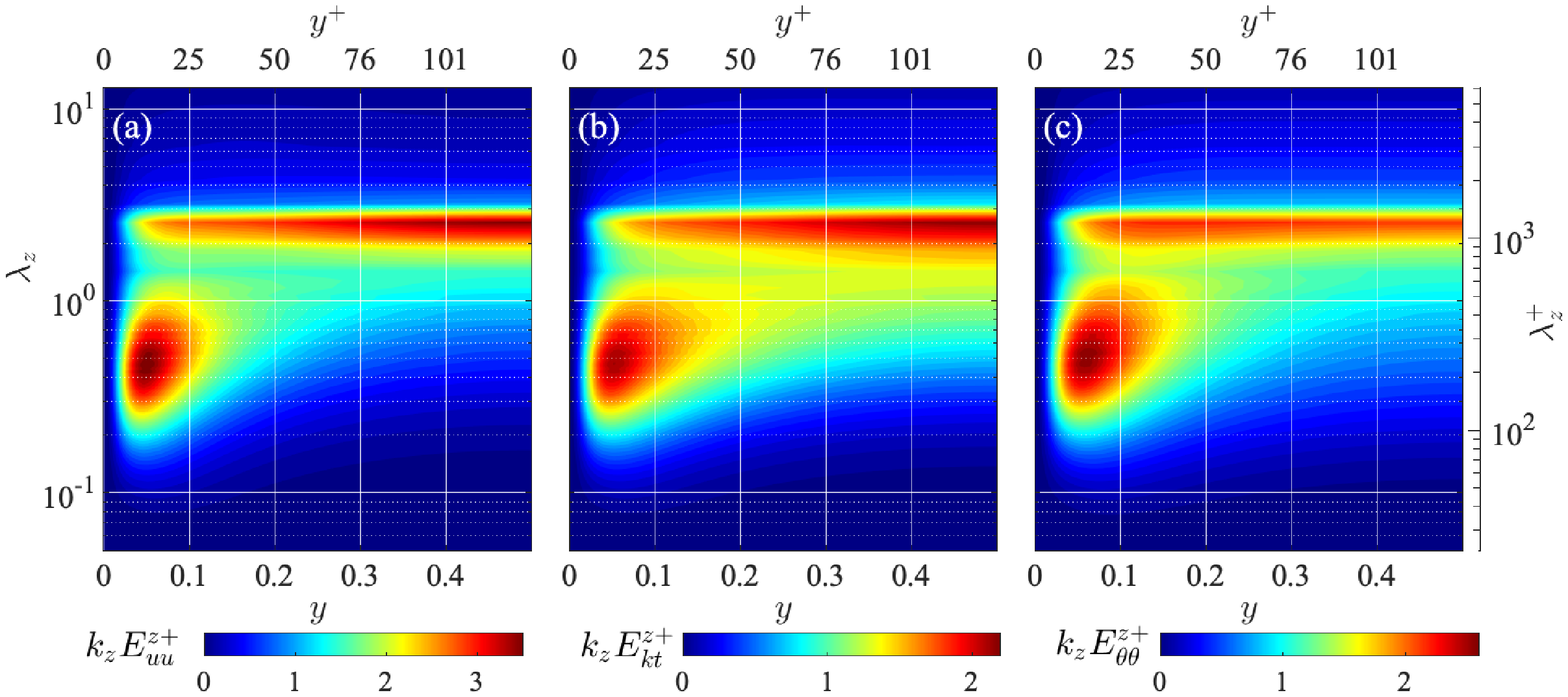}
    \caption{Distributions of spanwise one-dimensional spectra of (a)~streamwise turbulent energy $E^z_{uu}$, (b)~turbulent kinetic energy $E^z_{kt}$, and (c)~temperature fluctuation $E^z_{\theta \theta}$, presented in the same manner as in Fig.~\ref{fig:ex}.}
    \label{fig:ez}
\end{figure*}

Next, we focus on the spectral quantities of the turbulent statistics. Figure~\ref{fig:ex} presents the space-wavelength ($y$-$\lambda_x$) diagrams of premultiplied one-dimensional streamwise energy spectra, comparing those of the streamwise turbulent energy $E^x_{uu}$, the turbulent kinetic energy $E^x_{kt}$ (here, $k_t=(\ave{u^2}+\ave{v^2}+\ave{w^2})$/2), and the temperature fluctuation $E^x_{\theta\theta}$. Note here that these spectra distributions are also shown only for the lower half of the channel $0\ll y \ll 0.5$, as they are averaged between the upper and lower halves of the channel. As shown in the panel~(a), the distribution of $E^x_{uu}$ indicates an energy peak in the near-wall region at relatively small streamwise wavelength $\lambda_x^+ \approx 600$, which corresponds to the near-wall coherent structures. It is also observed that the spectral energy increases at the largest wavelength $\lambda_x=24$ throughout the channel, representing the very-large-scale structure. 

The energy spectrum of the temperature fluctuation~$E^x_{\theta \theta}$ given in Fig.~\ref{fig:ex}(c) presents a qualitatively similar distribution to that of $E^x_{uu}$, showing the energy peaks at relatively small wavelength near the wall and at the largest wavelength $\lambda_x =24$ in the channel-core region. This observation is consistent with the fact that the profiles of $\ave{u^2}$ and $\ave{\theta^2}$ are similar to each other. One can, however, also find that the distributions of the temperature-fluctuation spectrum $E^x_{\theta \theta}$ spans to smaller wavelengths than that of the streamwise turbulent energy $E^x_{uu}$: while in the $E^x_{uu}$ distribution the energy larger than 30\% of the peak magnitude is distributed roughly in the range $\lambda_x > 0.5$, the energy with the same magnitude spans down to about $\lambda_x = 0.2$ in the $E^x_{\theta\theta}$ distribution. The difference between the distributions of $E^x_{uu}$ and $E^x_{\theta \theta}$ is found particularly clearly in the channel-core region, as the energy peak of $E^x_{uu}$ is located at around $\lambda_x \approx 3$, while that of the $E^x_{\theta \theta}$ distribution is at $\lambda_x \approx 1$. This observation is consistent with the report by Hane~{\it et al.}\cite{hane_2006}.

The difference between the distributions of $E^x_{uu}$ and $E^x_{\theta \theta}$ is partly attributable to the energy redistribution by the pressure-strain correlation, as it absorbs a certain amount of energy from $E^x_{uu}$ and transfers it to other components $E^x_{vv}$ and $E^x_{ww}$ at the relatively small wavelengths\cite{lee_2019,kawata_2021}. Figure~\ref{fig:ex}(b) presents the turbulent kinetic energy spectrum $E^x_{kt}=(E^x_{uu}+E^x_{vv}+E^x_{ww})/2$, and it is seen here that the distribution of $E^x_{kt}$ is slightly shifted to small-wavelength side compared to that of $E^x_{uu}$ alone, because of the spectral energy at relatively small $\lambda_x$ possessed by $E^x_{vv}$ and $E^x_{ww}$. However, the temperature fluctuation spectrum $E^x_{\theta \theta}$ is still distributed in somewhat smaller wavelength range than $E^x_{kt}$, compare Figs.~\ref{fig:ex}(b) and (c). 

This shift of the distribution of $E^x_{\theta \theta}$ towards small $\lambda_x$ side in comparison with the $E^x_{kt}$ distribution might be a somewhat surprising outcome, because the effect of molecular diffusion is more significant in the temperature field than in the velocity field as the Prandtl number is smaller than 1. Hence, one might expect that small-scale fluctuations would be damped in the fluctuating temperature field as compared to the velocity field. It has, however, turned out that small-scale fluctuations are more energetic in the temperature field than in the velocity field, as observed in Fig.~\ref{fig:ex}.

Such difference between the spectra of the velocity and temperature fluctuations is not observed from the spanwise spectra. Figure~\ref{fig:ez} presents the distributions of the spanwise one-dimensional spectra $E^z_{uu}$, $E^z_{kt}$, and $E^z_{\theta \theta}$ in the same manner as in Fig.~\ref{fig:ex}. As shown in the panel~(a), the distribution of the streamwise turbulent energy spectrum $E^z_{uu}$ shows a near-wall energy peak at $\lambda_z^+ \approx 100$ and a significant energy band across the channel at $\lambda_z \approx 2.5$, which correspond to the near-wall coherent and the very-large-scale structures, respectively. The temperature-fluctuation spectrum $E^z_{\theta \theta}$ presents a similar distribution to $E^z_{uu}$ in terms of the peak wavelengths, as shown in Fig.~\ref{fig:ez}(b). As observed here, the spectral energies $E^z_{uu}$ and $E^z_{\theta \theta}$ are distributed in the same $\lambda_z$ range, which is in contrast to $E^x_{uu}$ and $E^x_{\theta \theta}$ as a function of the streamwise wavelength $\lambda_x$. 

It is also noteworthy here that the turbulent kinetic energy spectrum $E^z_{kt}$ also gives a similar distribution to $E^z_{uu}$, as shown in Fig.~\ref{fig:ez}(b), unlike the streamwise counterparts of $E^x_{uu}$ and $E^x_{kt}$. This is presumably because the $\lambda_z$ range where the lateral turbulent energies $E^z_{vv}$ and $E^z_{ww}$ are distributed are not so small as compared to the streamwise turbulent energy $E^z_{uu}$. 

\begin{figure}
    \centering
    \includegraphics[width=1\hsize]{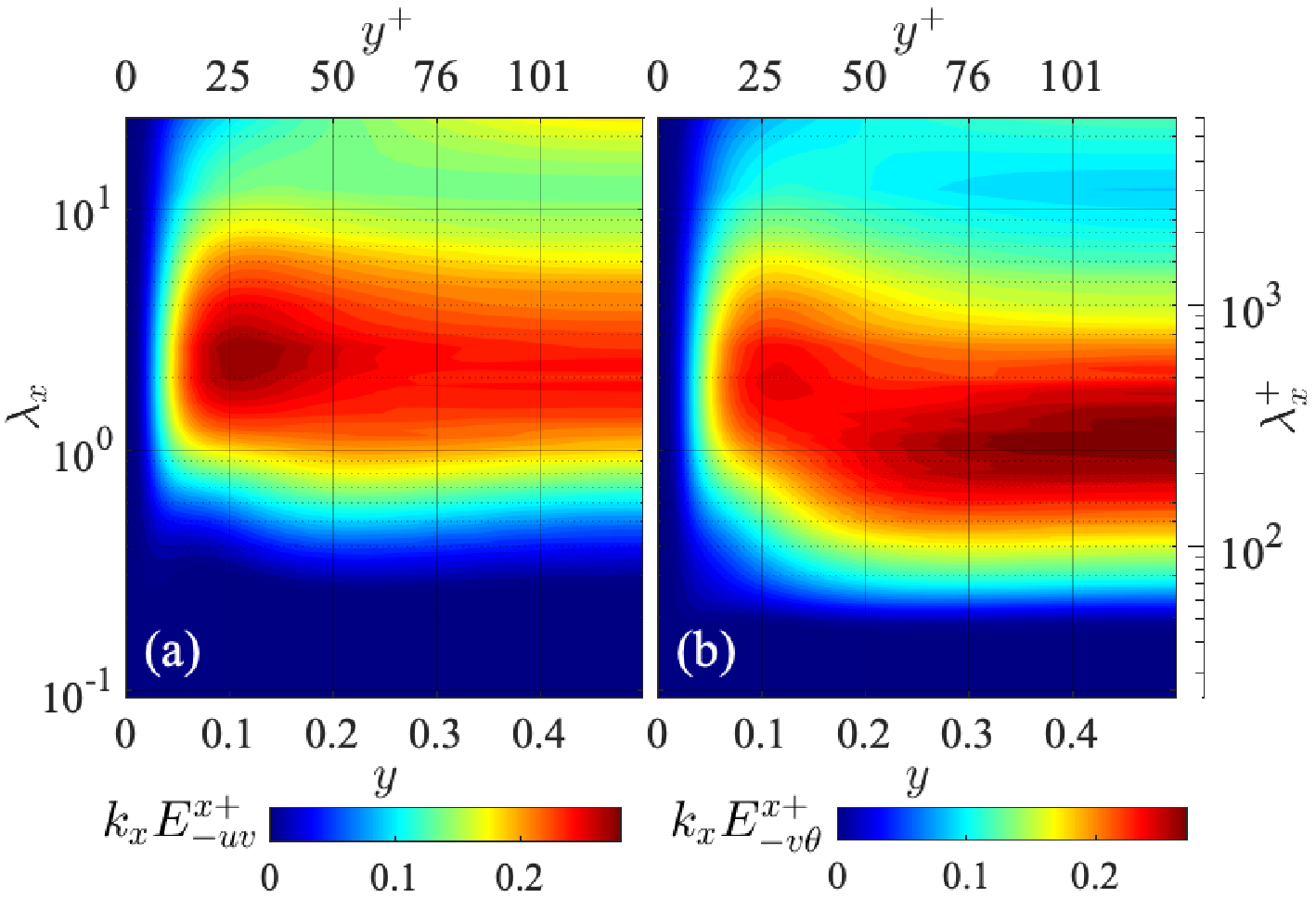}
    \caption{Distributions of streamwise one-dimensional cospectra of (a) the Reynolds shear stress $k_x E^x_{-uv}$ and (b) turbulent heat flux $k_x E^x_{-v\theta}$, presented in the same manner as in Fig.~\ref{fig:ex}. The values are scaled by $u_\tau^{\ast 2}$ and $u_\tau^\ast T_\tau^\ast$, respectively.}
    \label{fig:uv_vtx}
    \vspace{1em}
    \includegraphics[width=1\hsize]{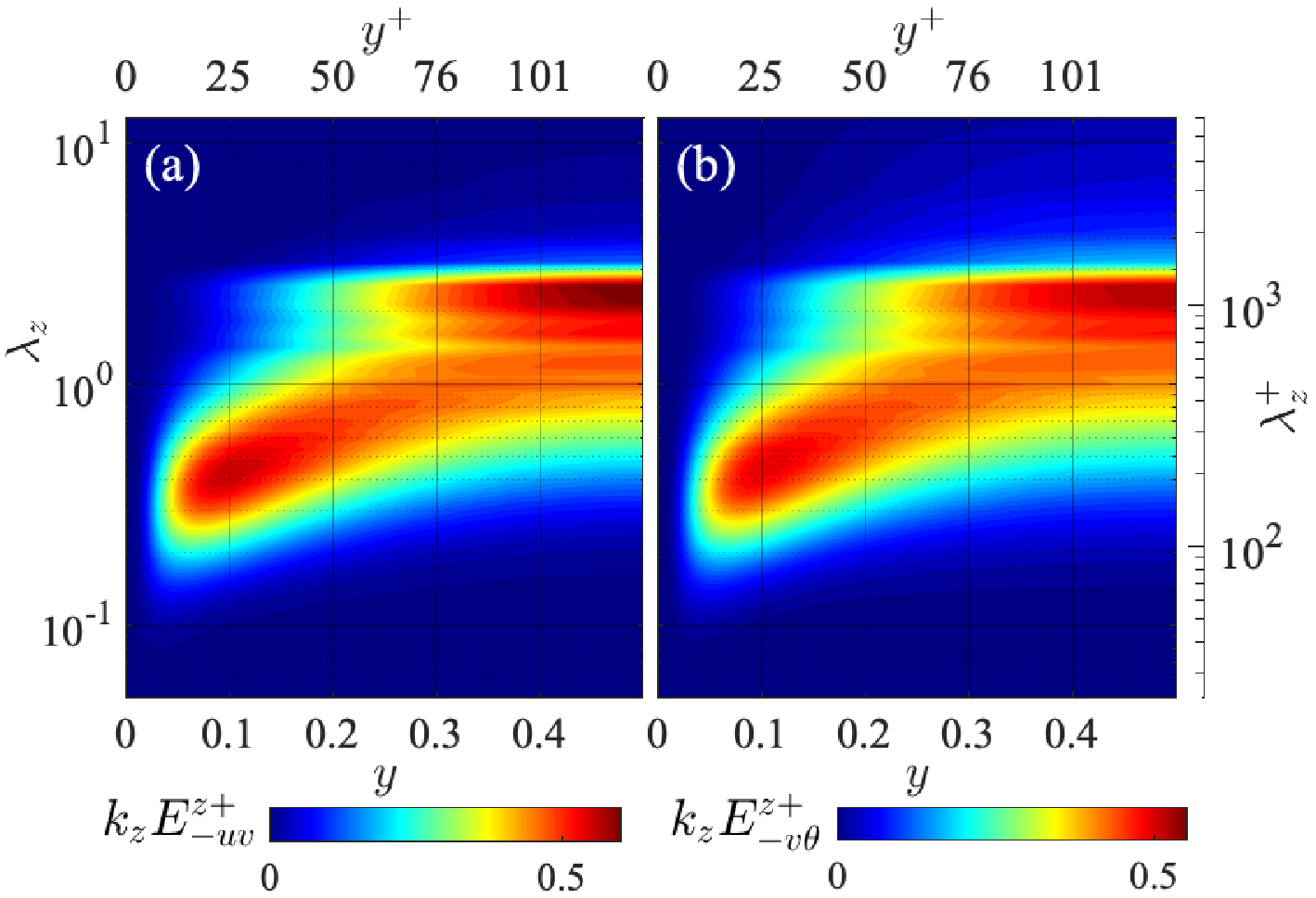}
    \caption{Spanwise one-dimensional spectra of (a) the Reynolds shear stress $E^z_{-uv}$ and (b) turbulent heat flux $E^z_{-v\theta}$, presented in the same manner as in Fig.~\ref{fig:uv_vtx}. }
    \label{fig:uv_vtz}
\end{figure}

Similar tendencies are also observed by the spectra of the turbulent momentum and heat fluxes. Figures~\ref{fig:uv_vtx} and \ref{fig:uv_vtz} present the distributions of one-dimensional streamwise and spanwise cospectra of turbulent fluxes, respectively, comparing those of the Reynolds shear stress and the turbulent heat flux.  Note here that as $-\ave{uv}$ and $-\ave{v\theta}$ represent the wall-normal transports of momentum and heat by turbulent fluid motions, respectively, their cospectra given in Figs.~\ref{fig:uv_vtx} and \ref{fig:uv_vtz} represent the turbulent momentum and heat transfers at each length scale. As shown in Fig.~\ref{fig:uv_vtx}, the distributions of the streamwise spectra of the Reynolds shear stress $E^x_{-uv}$ and the turbulent heat flux $E^x_{-v\theta}$ are qualitatively similar, but the distribution of $E^x_{-v\theta}$ is somewhat shifted towards the smaller-wavelength side as compared to $E^x_{-uv}$ particularly in the channel-core region. The spanwise spectra $E^z_{-uv}$ and $E^z_{-v\theta}$, on the other hand, give quite similar distributions as presented in Fig.~\ref{fig:uv_vtz}, and the difference between the spanwise length scales are not observed, consistently to the tendencies shown by the turbulent energy spectra in Figs.~\ref{fig:ex} and \ref{fig:ez}. These observations indicate that the turbulent momentum and heat transfers are certainly similar scale-by-scale, but the turbulent heat flux tends to occur at smaller streamwise length scales. 

The Reynolds shear stress $-\ave{uv}$ and the turbulent heat flux $-\ave{v\theta}$ are closely related to the wall shear stress $\tau_\mathrm{w}$ and the wall heat flux $q_\mathrm{w}$, respectively. Integrating the averaged momentum and energy transport equations across the chanel, one obtains 
\begin{subequations} \label{eq:fik}
\begin{align}
    \tau_\mathrm{w} &= \underbrace{\frac{1}{Re_\mathrm{w}}}_{\tau_\mathrm{w,lam}} + \underbrace{\int^1_0 (-\ave{uv}) \mathrm{d}y}_{\Delta \tau_\mathrm{w}}, \\
% \end{align}
% \begin{align}
    q_\mathrm{w} &= \underbrace{\frac{1}{Pr Re_\mathrm{w}}}_{q_\mathrm{w,lam}} + \underbrace{\int^1_0 (-\ave{v\theta}) \mathrm{d}y}_{\Delta q_\mathrm{w}},
\end{align}
\end{subequations}
where the first terms of $\tau_\mathrm{w,lam}$ and $q_\mathrm{w,lam}$ are the laminar values of the shear stress and the heat flux on the wall, respectively, and the second terms represent the increase from them due to turbulent fluid motions. Equations~(\ref{eq:fik}) are the Fukagata-Iwamoto-Kasagi identity\citep{fukagata_2002} for the plane Couette configuration. Taking the streamwise one-dimensional spectra $E^x_{-uv}$ and $E^x_{-v\theta}$ for example, $\Delta \tau_\mathrm{w}$ and $\Delta q_\mathrm{w}$ are expressed in terms of $E^x_{-uv}$ and $E^x_{-v\theta}$ as
\begin{subequations} \label{eq:dtau_dqw}
\begin{align}
    \Delta \tau_\mathrm{w} &= \int^1_0 \int^\infty_0 E^x_{-uv} (y,k_x)\mathrm{d} k_x \mathrm{d}y 
    % &= \int^\infty_0 \left( \int^1_0 E^x_{-uv} (y,k_x)  \mathrm{d}y\right) \mathrm{d} k_x \nonumber\\
     =\int^\infty_0 \mathcal{E}^x_{-uv}(k_x) \mathrm{d}k_x, \\
    \Delta q_\mathrm{w} &= \int^1_0 \int^\infty_0 E^x_{-v\theta}(y,k_x) \mathrm{d} k_x \mathrm{d}y 
    % &= \int^\infty_0 \left( \int^1_0 E^x_{-v\theta}(y,k_x) \mathrm{d}y\right) \mathrm{d} k_x \nonumber \\
    = \int^\infty_0 \mathcal{E}^x_{-v\theta} (k_x) \mathrm{d}k_x,
\end{align}
where $\mathcal{E}^x_{-uv}$ and $\mathcal{E}^x_{-v\theta}$ are the spectra of turbulent momentum and heat transfers
integrated across the channel:
\begin{align}
    \mathcal{E}^x_{-uv}=\int^1_0 E^x_{-uv} (y,k_x) \mathrm{d}y, \\% \quad
    \mathcal{E}^x_{-v\theta}=\int^1_0 E^x_{-v\theta} (y,k_x) \mathrm{d}y.
\end{align}
\end{subequations}
Equations (\ref{eq:dtau_dqw}) show that the integrated spectra $\mathcal{E}^x_{-uv}$ and $\mathcal{E}^x_{-v\theta}$ represent the spectral contents of $\Delta \tau_\mathrm{w}$ and $\Delta q_\mathrm{w}$, respectively. Similarly, one can also similarly define the integrated spanwise spectra, $\mathcal{E}^z_{-uv}(k_z)$ and $\mathcal{E}^z_{-v\theta}(k_z)$, which represent the contributions to $\Delta \tau_\mathrm{w}$ and $\Delta q_\mathrm{w}$ from each spanwise wavenumber.

\begin{figure}
    \centering
    \includegraphics[width=0.85\hsize]{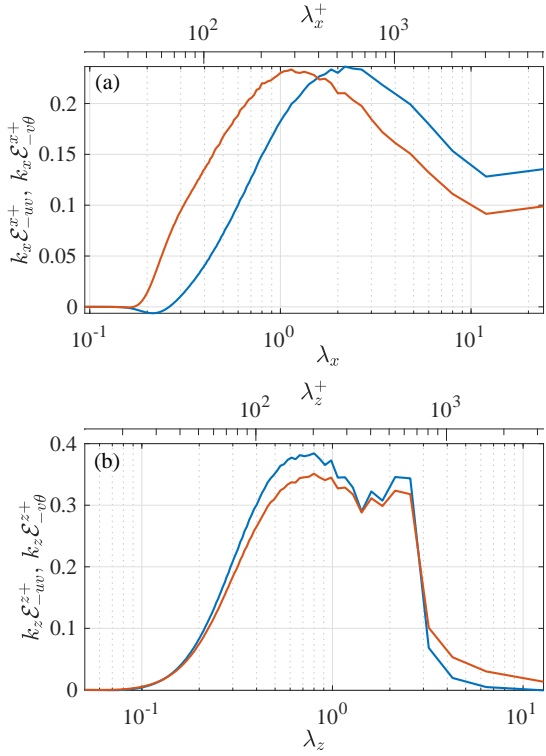}
    \caption{Premultiplied spectra of turbulent wall shear stress and turbulent wall heat flux: (a) streamwise spectra $k_x \mathcal{E}^x_{-uv}$ and $k_x \mathcal{E}^x_{-v\theta}$, (b) spanwise spectra $k_z \mathcal{E}^z_{-uv}$ and $k_z \mathcal{E}^z_{-v\theta}$. The values are scaled by $u_\tau^{\ast 2} h^\ast$ and $u_\tau^\ast T_\tau^\ast h^\ast$, respectively. }
    \label{fig:EUV_VT}
\end{figure}

The spectra of the wall shear stress and the wall heat flux are presented in Fig.~\ref{fig:EUV_VT}. In the panel~(a), the streamwise one-dimensional spectra of the wall shear stress $\mathcal{E}^x_{-uv}$ and that of the wall heat flux $\mathcal{E}^x_{-v\theta}$ are compared. One can see here that $\mathcal{E}^x_{-uv}$ and $\mathcal{E}^x_{-v\theta}$ show similar profiles to each other. Both profiles have a small peak at the largest streamwise wavelength $\lambda_x=24$, which represents the contribution by the very-large-scale structures to $\Delta \tau_\mathrm{w}$ and $\Delta q_\mathrm{w}$, and also show their largest peak at relatively small wavelengths on the order of $\lambda_x \approx 1$. However, the peak of the $\mathcal{E}^x_{-v\theta}$ profile is located at a somewhat smaller wavelength than that of the $\mathcal{E}^x_{-uv}$, which corresponds to the difference between the distributions of $E^x_{-uv}$ and $E^x_{-v\theta}$ observed in Fig.~\ref{fig:uv_vtx}. This means, again, that the streamwise length scales that mainly contribute to the wall heat flux is somewhat smaller to those responsible to the wall shear stress, despite $Pr<1$. 
The spanwise spectra of the wall shear stress $\mathcal{E}^z_{-uv}$ and the wall heat flux $\mathcal{E}^z_{-v\theta}$, however, does not show such difference in the scales, as presented in Fig.~\ref{fig:EUV_VT}(b). Both profiles of $\mathcal{E}^z_{-uv}$ and $\mathcal{E}^z_{-v\theta}$ indicate significant peaks at $\lambda_z^+ \approx 100$ and $\lambda_z = 2.1$--2.5, corresponding to the near-wall and very-large-scale structures, respectively, and these peaks of $\mathcal{E}^z_{-uv}$ and $\mathcal{E}^z_{-v\theta}$ are located at the same wavelengths, unlike the peaks observed in the streamwise spectra $\mathcal{E}^x_{-uv}$ and $\mathcal{E}^x_{-v\theta}$. 

\subsection{Interscale and Spatial Fluxes of the Reynolds Stresses and Temperature-Related Statistics}

Now we focus on the interscale and spatial fluxes of the temperature-related statistics. As shown in the previous section, the streamwise spectra indicate that the temperature fluctuation is more energetic at small scales than the velocity fluctuation, although the spanwise spectra do not show any significant difference. Hence, we focus on the transports of the streamwise spectra in the following. 

\begin{figure}
\centering
    \includegraphics[width=1\hsize]{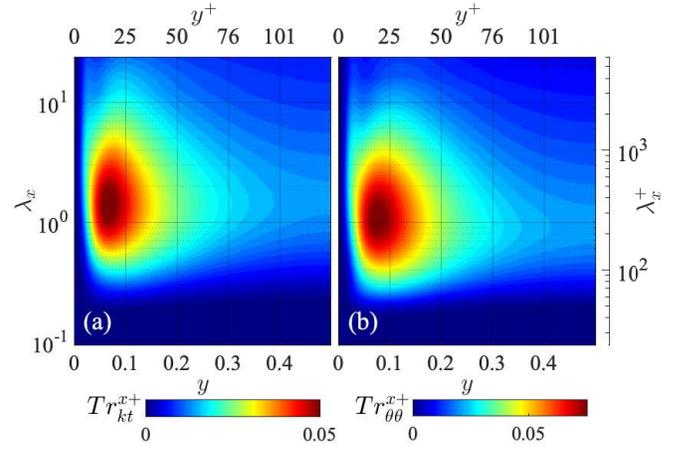}
    \caption{Space-wavelength ($y$-$\lambda_x$) diagrams of streamwise interscale fluxes of (a) turbulent kinetic energy~$Tr^x_{kt}$ and (b) temperature fluctuation $Tr^x_{\theta\theta}$. The values are scaled by $u_\tau^{\ast 4}/\nu^\ast$ and $u_\tau^{\ast  2} T_\tau^{\ast 2}/\nu^\ast$, respectively.}
    \label{fig:trxuu_tt}
\end{figure}

Figure~\ref{fig:trxuu_tt} compares the distributions of the interscale fluxes (in the streamwise wavenumber direction) of the turbulent kinetic energy $Tr^x_{kt}$ and the temperature fluctuation $Tr^x_{\theta \theta}$. It should be noted here that according to the definition in Eq.~(\ref{eq:rstels}) the positive values of the interscale flux represent the transfer from larger to smaller scales. As shown in Fig.~\ref{fig:trxuu_tt}(a), the interscale turbulent energy flux $Tr^x_{kt}$ basically indicates forward energy transfers, i.e., from larger to smaller scales. 
\begin{figure}
\centering
    \includegraphics[width=1\hsize]{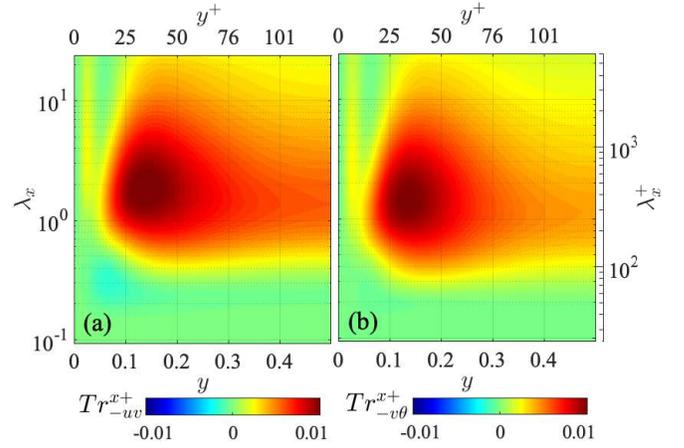}    
    \caption{Streamwise interscale fluxes of (a) the Reynolds shear stress~$Tr^x_{-uv}$ and (b) turbulent heat flux $Tr^x_{-v \theta}$, presented in the same manner as in Fig.~\ref{fig:trxuu_tt}.}
\label{fig:trxuv_vt} 
\end{figure}
The distribution of the interscale flux of the temperature fluctuation $Tr^x_{\theta \theta}$ is similar to that of the $Tr^x_{kt}$ distribution as presented in Fig.~\ref{fig:trxuu_tt}(b), but it appears to be somewhat shifted towards the smaller-wavelength side. The difference is particularly visible in the central region of the channel. This tendency of the $Tr^x_{\theta \theta}$ distribution is consistent with the observation in Fig.~\ref{fig:ex} that the distribution of $E^x_{\theta \theta}$ spans to smaller wavelengths than the $E^x_{kt}$ distribution. 

Figure \ref{fig:trxuv_vt} compares the distributions of the streamwise interscale fluxes of the Reynolds shear stress $Tr^x_{-uv}$ and the temperature-velocity correlation  $Tr^x_{-v\theta}$. As shown, the distributions of $Tr^x_{-uv}$ and $Tr^x_{-v\theta}$ are similar to each other, similarly to the interscale fluxes of turbulent energies in Fig.~\ref{fig:trxuu_tt}. Both $Tr^x_{-uv}$ and $Tr^x_{-v\theta}$ basically present forward interscale transfers and their maxima are found at similar locations in the diagrams. It is shown, again, that the distribution of $Tr^x_{-v\theta}$ spreads to smaller wavelengths than the Reynolds shear stress flux $Tr^x_{-uv}$, which is consistent with the corresponding cospectrum distributions given in Fig.~\ref{fig:uv_vtx}. 

\begin{figure}
\centering
    \includegraphics[width=1\hsize]{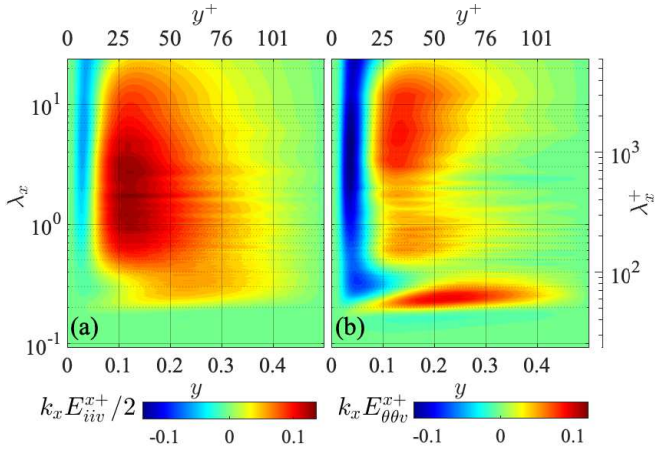}
    \caption{Space-wavelength ($y$-$\lambda_x$) diagram of premultiplied streamwise one-dimensional spectra of turbulent spatial fluxes: (a) turbulent kinetic energy transport $k_x E^x_{iiv}/2 = k_x(E^x_{uuv} +E^x_{vvv} +E^x_{wwv})/2$ and (b) temperature fluctuation transport $k_x E^x_{\theta \theta v}$. The values are scaled by $u_\tau^{\ast 3}$ and $u_\tau^\ast T_\tau^{\ast 2}$, respectively.}
    \label{fig:Exuuv_ttv}
\end{figure}

Next we investigate the spectra of turbulent spatial fluxes of the turbulent energies and the temperature-related statistics. Figure~\ref{fig:Exuuv_ttv} presents the distributions of the streamwise one-dimensional spectrum of the turbulent spatial transport of the kinetic energy $E^x_{iiv}/2=(E^x_{uuv}+E^x_{vvv}+E^x_{wwv})/2$ and that of the temperature fluctuation $E^x_{\theta \theta v}$, which represent the spatial transports of $k_t$ and $\ave{\theta^2}$ in $y$-direction at each wavelength $\lambda_x$, respectively: see also Eqs.~(\ref{eq:eijk}b) and (\ref{eq:si_tt}c) for their definitions. As shown in the panel~(a), the spatial energy transport spectrum $E^x_{iiv}/2$ indicates a broad energy distribution over a wide range of $\lambda_x$. The region of negative $E^x_{iiv}/2$ (where the turbulent kinetic energy is transported towards the wall) in the near-wall region spans from the largest wavelength ($\lambda_x=24.0$) to the middle wavelength range around $\lambda_x=1.0$ ($\lambda_x \approx 250$), while the positive region ($k_t$ is transported away from the bottom wall) spreads down to further smaller wavelengths around $\lambda_x = 0.2$ ($\lambda_x^+ \approx 50$). 

The distribution of the $\ave{\theta^2}$ transport spectrum, $E^x_{\theta \theta v}$, also presents contributions from a wide range of the streamwise wavelength $\lambda_x$ similarly to the $E^x_{iiv}/2$ distribution, as shown in Fig.~\ref{fig:Exuuv_ttv}(b). However, in the $E^x_{\theta \theta v}$ distribution, the negative spatial transport in the near-wall region is shown to occur at much smaller wavelengths than in the $E^x_{iiv}/2$ spectrum distribution, and there is a secondary peak of the positive spectral transport at small wavelength around $\lambda_x \approx 0.3$ ($\lambda^+ \approx 60$). This secondary peak accounts for about 13\% of the maximum level of the overall turbulent spatial transport of the temperature fluctuation~$\ave{\theta^2 v}$, while the turbulent energy transport spectrum $E^x_{iiv}/2$ in the same $\lambda_x$ range accounts for about 7\% of $(\ave{u^2v}+\ave{v^3}+\ave{w^2v})/2$. This difference indicates that the turbulent spatial transport of the temperature fluctuation is more significant at relatively small wavelengths compared to the turbulent kinetic energy transport. 
% The streamwise spectra of the spatial transport of the Reynolds shear stress and the turbulent heat flux, $E^x_{-uvv}$ and $E^x_{-\theta vv}$, present basically the same tendencies (not shown).

\begin{figure}
\centering
    \includegraphics[width=1\hsize]{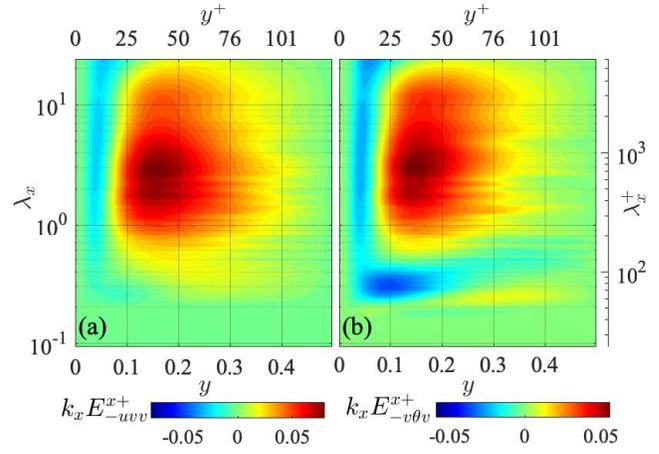}
    \caption{Streamwise one-dimensional spectra of turbulent spatial fluxes of (a) the Reynolds shear stress $k_x E^x_{-uvv}$ and (b) turbulent heat flux $k_x E^x_{-v \theta v}$. The values are scaled by $u_\tau^{\ast 3}$ and $u_\tau^{\ast 2} T_\tau^\ast$, respectively.}
    \label{fig:Exuvv_vtv}
\end{figure}

The spectral spatial fluxes of the Reynolds shear stress $E^x_{-uvv}$ and the turbulent heat flux $E^x_{-v\theta v}$ are also presented in Fig.~\ref{fig:Exuvv_vtv}. As shown in the panel~(a), the distribution of $E^x_{-uvv}$ indicates the wall-ward (negative) transports in the near-wall region and the in-ward (positive) transports in the channel core for a wide range of the wavelength similarly to the turbulent kinetic energy flux $E^x_{iiv}/2$. It is also seen by comparing the distributions of $E^x_{-uvv}$ (Fig.~\ref{fig:Exuvv_vtv}a) and $E^x_{iiv}/2$ (Fig.~\ref{fig:Exuuv_ttv}a) that the Reynolds shear stress transport occurs at relatively large wavelength compared to the turbulent kinetic energy transport. The spectrum of the turbulent heat flux transport $E^x_{-v\theta v}$ also present a similar distribution, as shown in the panel~(b), where the negative and positive transports are indicated in the near- and far-wall regions, respectively, at relatively large wavelengths. It is also seen that the negative transports take place in a wide range of the channel at relatively small wavelengths $\lambda_x \approx 0.3$. Such spatial transports at small wavelength are also observed by the temperature-fluctuation flux $E^x_{\theta \theta v}$ as shown in Fig.~\ref{fig:Exuuv_ttv}(b), which is presumably related with the fact that the temperature fluctuation at small streamwise length scale is more energetic than the velocity fluctuation. 

As described so far, the spectral distributions of the fluctuating velocity and temperature fields have been investigated including their interscale and spatial transports, and it has been observed that the temperature-related statistics basically show similar distributions to the corresponding turbulence statistics, which indicates a close similarity between the fluctuating velocity and temperature fields. Some differences have, however, also been found. A particularly interesting observation is that the temperature fluctuation at small streamwise length scales is more energetic than those in the fluctuating velocity field, although $Pr<1$ and therefore the effect of molecular diffusion is greater in the temperature field. This point will be discussed in detail in the next section.

\section{Discussion} \label{sec:discussion}

\subsection{Spectral Analysis on Transport of Turbulent Kinetic Energy and Temperature Fluctuation}

Now we investigate the budget balance of the spectral transport equations of the temperature-related statistics, in order to address the difference in the dominating streamwise length scales of the heat and momentum transfers observed in the previous section. First, we focus on the spectral transport of the turbulent kinetic energy and the temperature fluctuation.

\begin{figure}
    \centering
    \includegraphics[width=1\hsize]{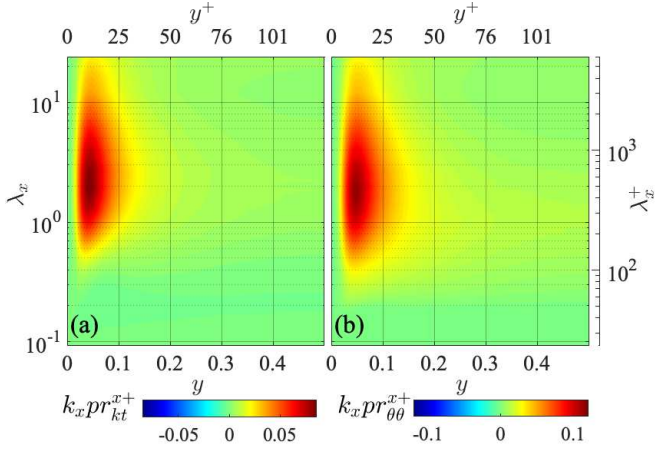}
    \caption{Space-wavelength ($y$-$\lambda_x$) diagrams of premultiplied streamwise one-dimensional spectra of (a) turbulent kinetic energy production $k_x pr^x_{uu}$ and (b) temperature fluctuation production $k_x pr^x_{\theta \theta}$. The values are scaled by $u_\tau^{\ast 4}/\nu^\ast$ and $u_\tau^{\ast 2} T_\tau^{\ast 2}/\nu^\ast$, respectively. }
    \label{fig:prxuu_tt}
\end{figure}

Figure~\ref{fig:prxuu_tt} compares the production spectra of the turbulent kinetic energy $pr^x_{kt}$ and the temperature fluctuation $pr^x_{\theta \theta}$. These production spectra are directly related to the Reynolds shear stress cospectrum $E^x_{-uv}$ and the turbulent heat flux spectrum $E^x_{-v\theta}$ as
\begin{subequations}
\begin{align}
    pr^x_{kt} &= E^x_{-uv} \frac{\mathrm{d} U}{\mathrm{d} y}, \\%\quad
    pr^x_{\theta \theta} &= 2 E^x_{-v\theta} \frac{\mathrm{d} \varTheta}{\mathrm{d} y},
\end{align}
\end{subequations}
in the present flow configuration, and therefore the $\lambda_x$ ranges where the productions of $k_t$ and $\ave{\theta^2}$ occur are determined by the distributions of $E^x_{-uv}$ and $E^x_{-v\theta}$, respectively. As shown in Fig.~\ref{fig:prxuu_tt}, these production spectra present similar distributions to each other, presenting a near-wall peak located at similar locations $y^+ \approx 16$ and $\lambda_x \approx 0.2$ ($\lambda_x^+ \approx 500$). However, corresponding to the distributions of $E^x_{-uv}$ and $E^x_{-v\theta}$ given in Fig.~\ref{fig:uv_vtx},  the distribution of the temperature-fluctuation production $pr^x_{\theta \theta}$ is shown to spread down to smaller wavelengths than the turbulent energy production $pr^x_{kt}$. The difference between the distributions of these production spectra is, however, not as distinct as that between the $E^x_{kt}$ and $E^x_{\theta \theta}$ distributions observed in Fig.~\ref{fig:ex}. 

\begin{figure}
    \centering
    \includegraphics[trim=0 57 0 0 ,clip,width=1\hsize]{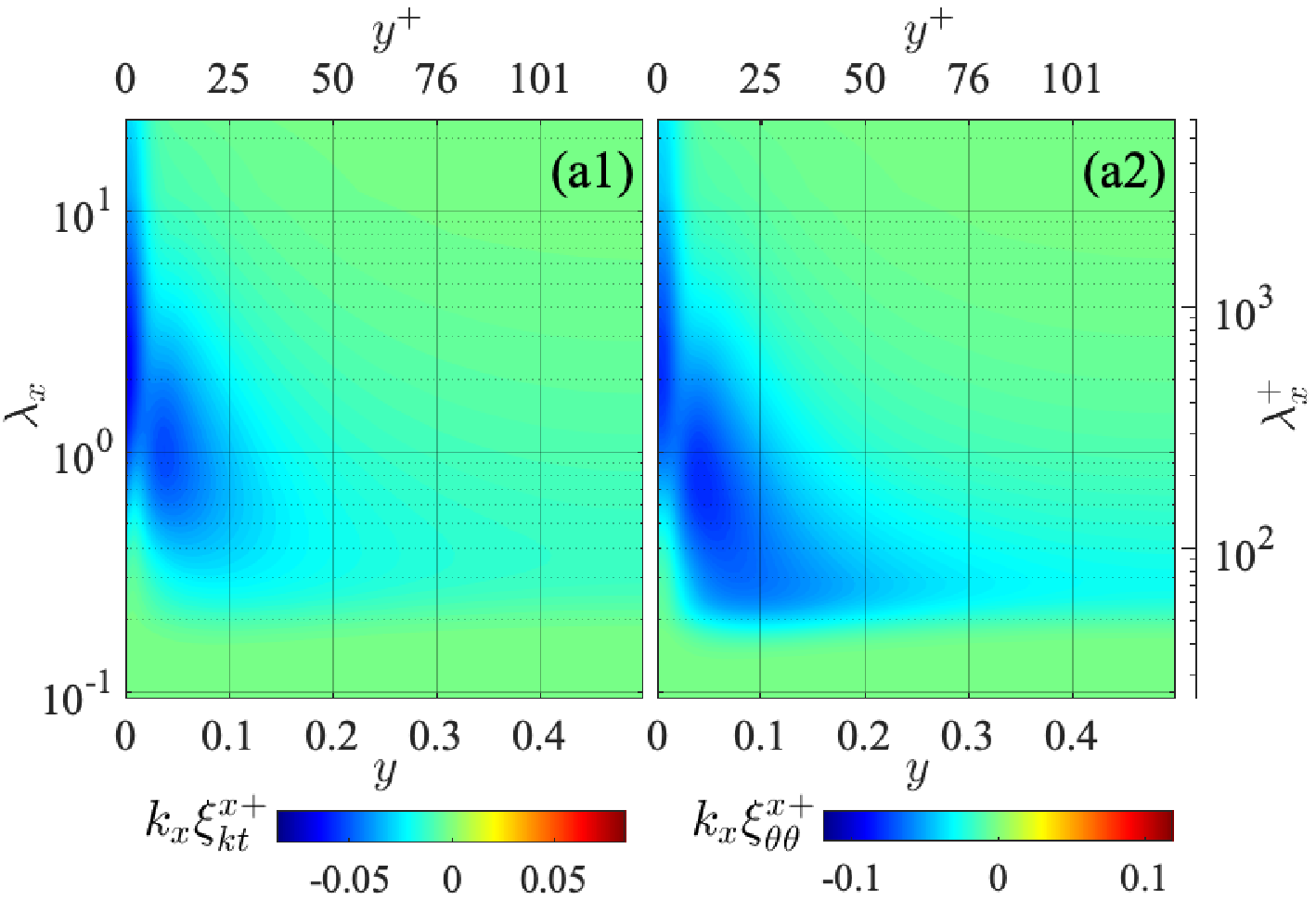}
    \includegraphics[trim=0 32 0 30 ,clip,width=1\hsize]{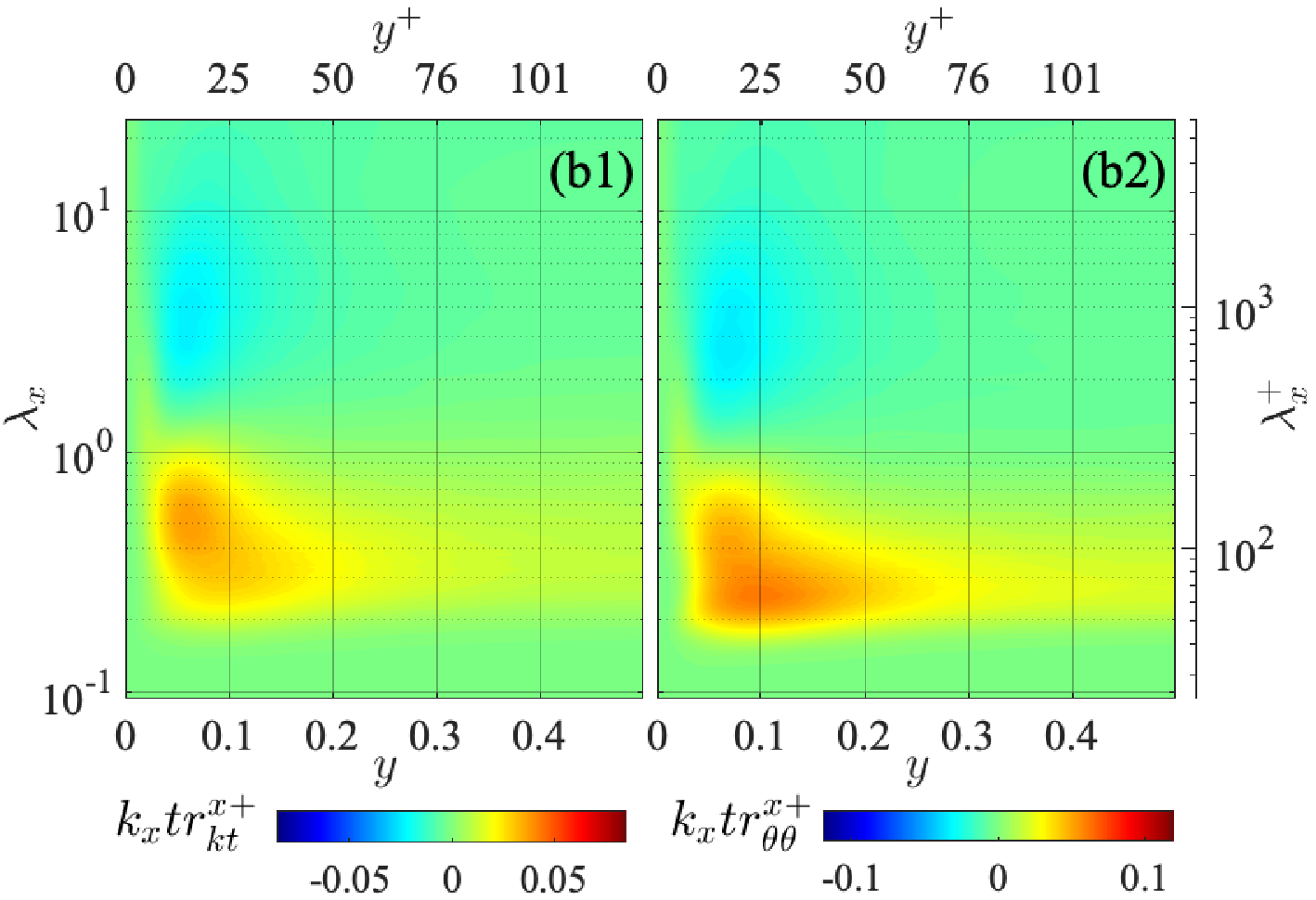}
        \caption{Distributions of viscous dissipation spectra (a1) $\xi^x_{kt}$ and $\xi^x_{\theta \theta}$ and interscale energy transports (b1) $tr^x_{kt}$ and (b2) $tr^x_{\theta \theta}$, presented in the same manner as in Fig.~\ref{fig:prxuu_tt}. The colour scales are also the same as in the corresponding panels of Fig.~\ref{fig:prxuu_tt}.}
    \label{fig:trx_xix}
\end{figure}

More clear differences in the transport budgets of $E^x_{kt}$ and $E^x_{\theta \theta}$ are found by investigating their viscous dissipations and interscale transports. Figures~\ref{fig:trx_xix}(a1) and (a2) present the distributions of the viscous dissipation spectra of the turbulent kinetic energy $\xi^x_{kt}$ and the temperature fluctuation $\xi^x_{\theta \theta}$, respectively. It is shown here that $\xi^x_{\theta\theta}$ at small wavelengths $\lambda_x$ is more significant than $\xi^x_{kt}$. The difference is particularly notable in the region away from the wall $y \geq 0.2$, where $\xi^x_{kt}$ is obviously weak as compared to the near-wall region while $\xi^x_{\theta\theta}$ indicates a certain energy dissipation at $\lambda_x \approx 0.25$ ($\lambda_x^+ \approx 70$) in the channel-core region. Such differences between the viscous dissipations are relatively distinct as compared to the production spectra $pr^x_{kt}$ and $pr^x_{\theta\theta}$ and consistent with the observations by Antonia \& Abe\cite{antonia_2009}. However, again, this might be a surprising observation since the effect of molecular diffusion is more significant in the temperature field as the Prandtl number $Pr$ is smaller than 1.

The interscale transport of the turbulent kinetic energy $tr^x_{kt}$ and the temperature fluctuation $tr^x_{\theta \theta}$ are compared in Figs.~\ref{fig:trx_xix}(b1) and (b2). Both distributions indicate that the energy is transferred basically from relatively large wavelengths roughly in the range $1 < \lambda_x < 10$ to smaller wavelengths in $\lambda_x < 1$. It is, however, also clearly seen that $tr^x_{\theta \theta}$ delivers the energy to smaller wavelengths than $tr^x_{kt}$; in the near-wall region the peak of the energy gain by $tr^x_{kt}$ is located about $\lambda_x = 0.54$ ($\lambda_x^+ = 137$) while in the $tr^x_{\theta \theta}$ distribution the near-wall peak of energy gain is located around $\lambda_x = 0.25$ ($\lambda_x^+=63$). The difference between the $tr^x_{kt}$ and $tr^x_{\theta \theta}$ is more distinct in the core region of the channel, as the energy gains by $tr^x_{\theta \theta}$ are still clearly observed around $\lambda_x \approx 0.25$ compared to that by the kinetic energy transport by $tr^x_{kt}$. Hence, it can be interpreted here that the different behaviours of the viscous dissipations $\xi^x_{kt}$ and $\xi^x_{\theta \theta}$ observed in Figs.~\ref{fig:trx_xix}(a1) and (a2) are the consequences of such different energy transports by $tr^x_{kt}$ and $tr^x_{\theta\theta}$.

\begin{figure}
    \centering
    \includegraphics[trim=23 28 20 0,clip,width=0.85\hsize]{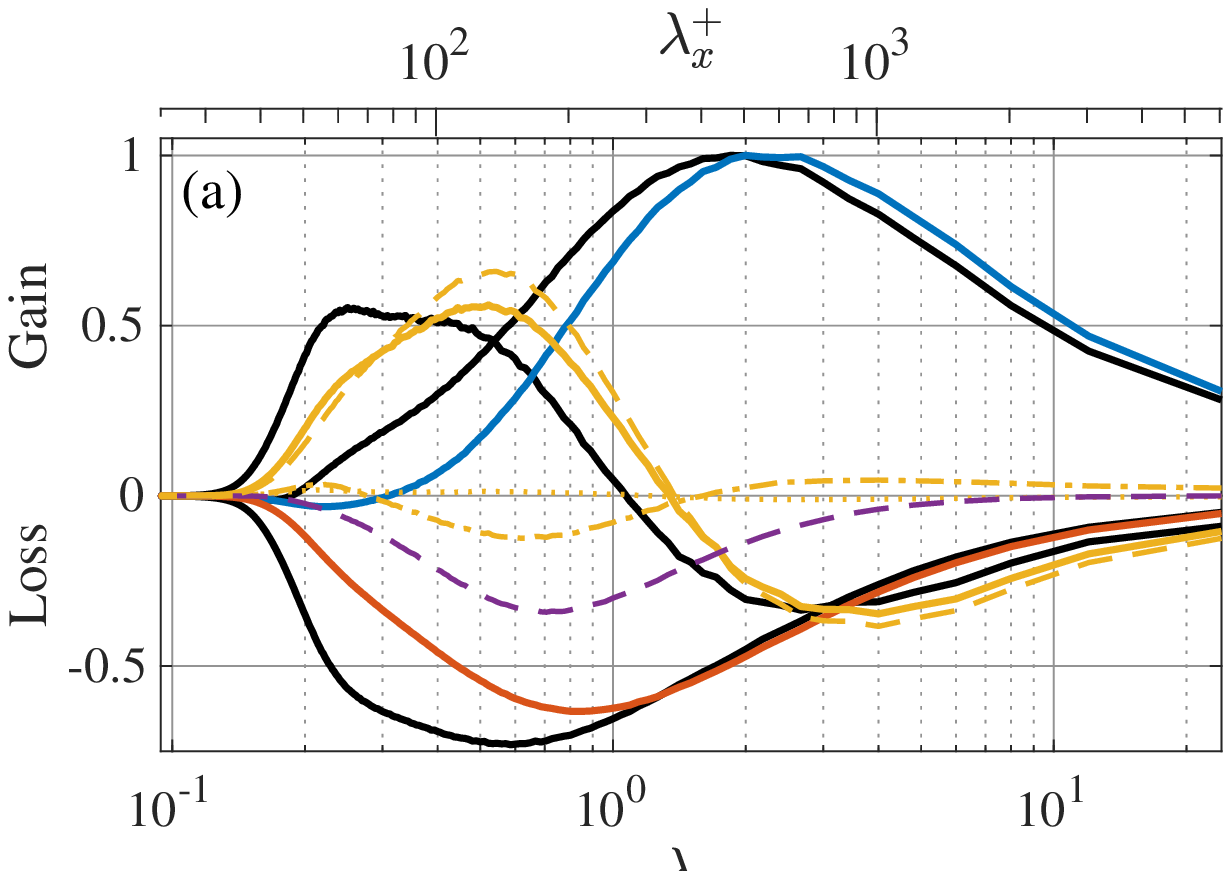}
    \includegraphics[trim=23 0 20 30,clip,width=0.85\hsize]{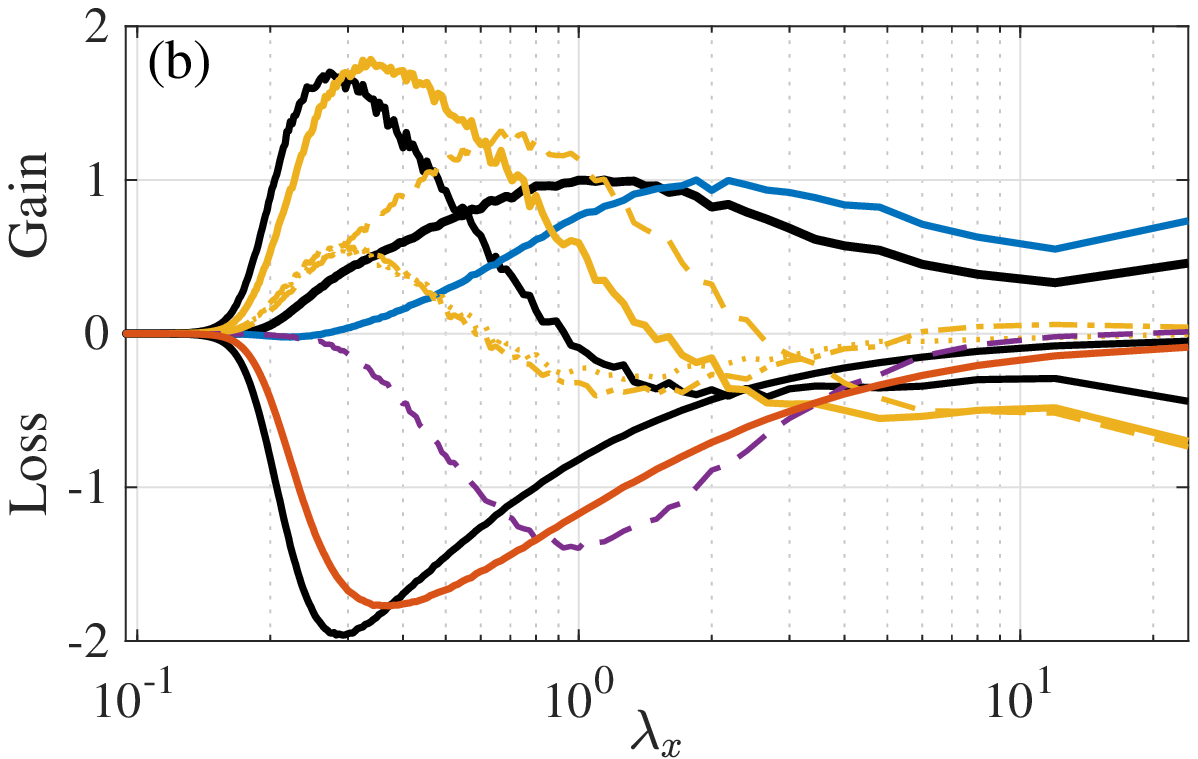}
    \caption{Comparison between transport budgets of $E^x_{kt}$ and $E^x_{\theta \theta}$ at (a)~$y^+=16$ and (b)~channel centre. The solid lines represent major terms in $E^x_{kt}$ transport, (blue) production $pr^x_{kt}$, (red)~viscous dissipation $\xi^x_{kt}$ and (yellow) interscale transport $tr^x_{tk}$, and the corresponding black solid lines are their counterparts in $E^x_{\theta \theta}$ transport. The yellow non-solid lines present different ingredients of $tr^x_{kt}$: `-- --', $tr^x_{uu}$; `$\cdots$', $tr^x_{vv}$; '-- $\cdot$ --', $tr^x_{ww}$. Pressure-strain cospectrum $\pi^x_{uu}$ is also shown as purple dashed lines. The values are scaled by the maximum value of the production spectra of each transport equation.}
    \label{fig:budget_nw_cc}
\end{figure}

Figure~\ref{fig:budget_nw_cc} presents further detailed balances between the production, viscous dissipation and interscale transport terms of the spectral transport equations, comparing those of the $E^x_{kt}$ and $E^x_{\theta \theta}$ equations at a near-wall location $y^+=16$ (a) and the channel centre (b). The values of each term are scaled by the maximum values of the corresponding production spectra. As shown in Fig.~\ref{fig:budget_nw_cc}, at both locations, $p^x_{kt}$ and $pr^x_{\theta\theta}$ have their peaks at similar wavelengths, $tr^x_{\theta \theta}$ delivers energy to smaller wavelengths than $tr^x_{kt}$, and $\xi^x_{\theta \theta}$ is more significant at smaller wavelengths than $\xi^x_{kt}$, as already described above. 

Among the different terms shown here, the difference between the interscale transport terms $tr^x_{kt}$ and $tr^x_{\theta \theta}$ at the near-wall location $y^+=16$ is particularly notable at small wavelengths, as shown in Fig.~\ref{fig:budget_nw_cc}(a). While the peak of energy gain by $tr^x_{kt}$ is located at $\lambda_x \approx 0.5$ ($\lambda_x^+\approx 130$), the profile of $tr^x_{\theta \theta}$ shows a plateau in the small wavelength range around $\lambda_x \approx 0.3$, and the maximum is located at $\lambda_x =0.25$ ($\lambda_x^+\approx 60$). Corresponding to such behaviours of the interscale transport terms, the viscous dissipation of the temperature fluctuation~$\xi^x_{\theta \theta}$ shows larger contributions at small wavelengths than the kinetic energy dissipation $\xi^x_{kt}$. 

At the channel centre, on the other hand, the difference between the interscale and viscous dissipation of $E^x_{kt}$ and $E^x_{\theta\theta}$ are not as distinct as in the near-wall location, as shown in Fig.~\ref{fig:budget_nw_cc}(b). However, the same tendency is still observed that the energy is transferred to and dissipated at relatively small wavelengths in the $E^x_{\theta \theta}$ transport as compared to the $E^x_{kt}$ transport.  

In each panel of Fig.~\ref{fig:budget_nw_cc}, the ingredients of the interscale kinetic energy transport $tr^x_{kt}$, i.e.~the transport of each velocity fluctuation components $tr^x_{uu}$, $tr^x_{vv}$ and $tr^x_{ww}$, are also presented for comparison. It is shown in the panel~(a) that at the near-wall location $y^+=16$ the interscale kinetic energy transport $tr^x_{kt}$ is basically dominated by the streamwise turbulent energy transport $tr^x_{uu}$, the transport of the wall-normal turbulent energy $tr^x_{vv}$ is negligibly small, and the transport of the spanwise turbulent energy~$tr^x_{ww}$ is relatively weak compared to $tr^x_{uu}$ but indicates both forward and inverse interscale energy transfers. The pressure-strain energy redistribution spectrum $\pi^x_{uu}$ is also presented here, and one can see that the profiles of $\pi^x_{uu}$ and $tr^x_{ww}$ have their negative peaks at roughly same wavelengths, which also correspond to the peak location of the energy gain by $tr^x_{uu}$. This indicates that the turbulent energy transferred from large $\lambda_x$ to the middle $\lambda_x$ range by $tr^x_{uu}$ is redistributed to $E^x_{ww}$ by $\pi^x_{uu}$, and then it is further transferred to both smaller and larger scales by $tr^x_{ww}$. 

In our previous study\cite{kawata_2021}, the role of such interscale transport of the spanwise turbulent energy $\ave{w^2}$ towards both larger and smaller wavelengths in the near-wall region was closely investigated. The energy transferred to smaller wavelengths by $tr^x_{ww}$ is dissipated by $\xi^x_{ww}$ in addition to the main energy dissipation by $\xi^x_{uu}$. On the other hand, the energy ``transferred back'' to larger wavelengths by $tr^x_{ww}$ is further transferred to the wall-normal turbulent energy $\ave{v^2}$ by the pressure-strain energy redistribution, and eventually leads to turbulent energy production at the large wavelengths through the Reynolds shear stress production. We suggested that such near-wall energy transports including reversed energy cascade by $tr^x_{ww}$ may correspond to the regeneration process of streamwise vortices in the self-sustaining cycle of the near-wall coherent structures, which is in agreement with the suggestion by Hamba\cite{hamba_2019}. 

At the channel centre, on the other hand, both $tr^x_{vv}$ and $tr^x_{ww}$ present contributions comparable to $tr^x_{uu}$ at small wavelengths around $\lambda_x \approx 0.3$, indicating forward energy cascades, as shown in Fig.~\ref{fig:budget_nw_cc}(b) (note here that the profiles of $tr^x_{vv}$ and $tr^x_{ww}$ shown in the panel are almost on top of each other). One can also see here that the energy gain by $tr^x_{uu}$ at middle wavelengths around $\lambda_x \approx 1$ is almost balanced by the energy sink by $\pi^x_{uu}$, and $tr^x_{vv}$ and $tr^x_{ww}$ are shown to transport energy from this $\lambda_x$ range to smaller wavelengths. Due to such contributions by $tr^x_{vv}$ and $tr^x_{ww}$, the contribution by the overall kinetic energy transport~$tr^x_{kt}$ is comparable to the temperature-fluctuation transport~$tr^x_{\theta \theta}$ at small wavelengths. This is in contrast to the transport balance in the near-wall region, where as shown in Fig.~\ref{fig:budget_nw_cc}(a) the energy transports by $tr^x_{vv}$ and $tr^x_{ww}$ are substantially small compared to $tr^x_{uu}$, and therefore the energy transport by $tr^x_{kt}$ is clearly inferior to $tr^x_{\theta \theta}$ at small wavelengths.

As described so far, the difference between the spectral transports of turbulent kinetic energy and temperature fluctuation is particularly found in the behaviours of the interscale energy transport terms and, therefore, we next look into further details of $tr^x_{kt}$ and $tr^x_{\theta \theta}$. According to Eqs.~(\ref{eq:Trij}) and (\ref{eq:si_tt}), the interscale energy fluxes $Tr^x_{kt}$ and $Tr^x_{\theta\theta}$ consist of six terms representing the effects of different velocity or temperature gradients:
\begin{subequations}\label{eq:dec_Tr}
\begin{align}
    Tr^x_{kt} = 
      \underbrace{-\ave{u_i^S u^S \pd{u_i^L}{x}}}_{Tr^{x,(1)}_{kt}}
      \underbrace{-\ave{u_i^S v^S \pd{u_i^L}{y}}}_{Tr^{x,(2)}_{kt}} 
      \underbrace{-\ave{u_i^S w^S \pd{u_i^L}{z}}}_{Tr^{x,(3)}_{kt}} \nonumber \\
      \underbrace{+\ave{u_i^L u^L \pd{u_i^S}{x}}}_{Tr^{x,(4)}_{kt}} 
      \underbrace{+\ave{u_i^L v^L \pd{u_i^S}{y}}}_{Tr^{x,(5)}_{kt}} 
      \underbrace{+\ave{u_i^L w^L \pd{u_i^S}{z}}}_{Tr^{x,(6)}_{kt}},
\end{align}
\begin{align}
    Tr^x_{\theta\theta} = 
       \underbrace{-2\ave{\theta^S u^S \pd{\theta^L}{x}}}_{Tr^{x,(1)}_{\theta\theta}} 
       \underbrace{-2\ave{\theta^S v^S \pd{\theta^L}{y}}}_{Tr^{x,(2)}_{\theta\theta}} 
       \underbrace{-2\ave{\theta^S w^S \pd{\theta^L}{z}}}_{Tr^{x,(3)}_{\theta\theta}} \nonumber \\
       \underbrace{+2\ave{\theta^L u^L \pd{\theta^S}{x}}}_{Tr^{x,(4)}_{\theta\theta}} 
       \underbrace{+2\ave{\theta^L v^L \pd{\theta^S}{y}}}_{Tr^{x,(5)}_{\theta\theta}} 
       \underbrace{+2\ave{\theta^L w^L \pd{\theta^S}{z}}}_{Tr^{x,(6)}_{\theta\theta}}.
\end{align}
\end{subequations}
Hence, the corresponding interscale transport terms in the equations of $E^x_{uu}$ and $E^x_{\theta \theta}$ are also decomposed accordingly: 
\begin{align}
    tr^x_{kt} = \sum^6_{i=1} tr^{x,(i)}_{kt}, \quad 
    tr^x_{\theta\theta} = \sum^6_{i=1} tr^{x,(i)}_{\theta \theta},
% 
    % tr^x_{uu} &= tr^{x,(1)}_{uu}+tr^{x,(2)}_{uu}+tr^{x,(3)}_{uu}
    %             +tr^{x,(4)}_{uu}+tr^{x,(5)}_{uu}+tr^{x,(6)}_{uu} \\
    % tr^x_{\theta\theta} &= tr^{x,(1)}_{\theta\theta}+tr^{x,(2)}_{\theta\theta}
    %                       +tr^{x,(3)}_{\theta\theta}+tr^{x,(4)}_{\theta\theta}
    %                       +tr^{x,(5)}_{\theta\theta}+tr^{x,(6)}_{\theta\theta} 
\end{align}
where $tr^{x,(i)}_{kt}$ and $tr^{x,(i)}_{\theta \theta}$ are the $k_x$-derivatives of the corresponding terms in Eqs.~(\ref{eq:dec_Tr}).  

\begin{figure}
    \centering
    \includegraphics[width=1\hsize]{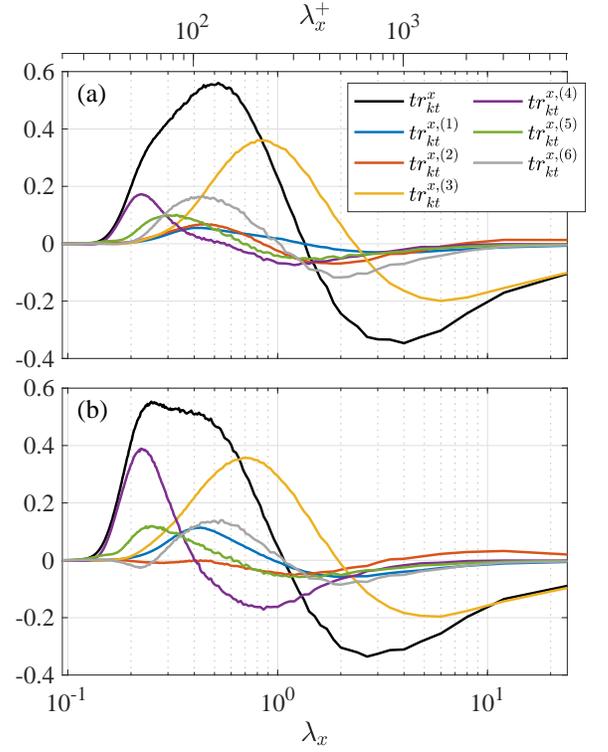}
    \caption{Comparisons of different terms in interscale energy transport terms (a) $tr^x_{uu}$ and (b) $tr^x_{\theta \theta}$ at near-wall location $y^+=16$. The values are scaled by the peak values of the corresponding production spectra. The colours of the plots in the panel~(b) presents the same terms as in the panel~(a).}
    \label{fig:trxi_nw}
\end{figure}

The various terms in the interscale transports $tr^x_{kt}$ and $tr^x_{\theta \theta}$ are compared for a near-wall location $y^+=16$ in Fig.~\ref{fig:trxi_nw}. As shown in the panel~(a), among various temrs in the turbulent kinetic energy transport the third term $tr^{x,(3)}_{kt}$ is shown to play particularly significant role at relatively large wavelengths, accounting for most of energy transfers from the largest wavelengths in the range $\lambda_x \approx 10$ ($\lambda_x^+ > 1000$) to the middle wavelengths around $\lambda_x \approx 1$ ($\lambda_x^+ \approx 200$). This term $tr^{x,(3)}_{kt}$ is associated with the spanwise gradient of the large-scale fluctuating velocity $\partial u_i^L/\partial z$, as can be seen in Eq.~(\ref{eq:dec_Tr}a). Therefore, the significant contribution of this term may represent the streak instabilities consisting of the self-sustaining cycle of the near-wall structures, as discussed in some earlier studies\cite{kawata_2021}. This may also be consistent with the observations by Wang {\it et al.}\cite{wang_2020} of forward scattering events associated with large-scale motions. 
The other terms of $tr^x_{kt}$ are shown to have relatively small contributions compared to the third term~$tr^{x,(3)}_{kt}$, indicating forward energy transfers at certainly smaller wavelengths. Among them, the fourth term $tr^{x,(4)}_{kt}$ provides a relatively distinct contributions, accounting for the energy supplies to smallest wavelengths $\lambda_x \approx 0.22$ ($\lambda_x^+\approx 50$). As shown by Eq.~(\ref{eq:dec_Tr}a), the fourth term $tr^{x,(4)}_{kt}$ is associated with the streamwise gradients of small-scale velocities $\partial u_i^S/\partial x$. 

Similar trends are also found in the behaviours of decomposed interscale transports of temperature fluctuation $tr^{x,(i)}_{\theta \theta}$. As presented in Fig.~\ref{fig:trxi_nw}(b), the third term~$tr^{x,(3)}_{\theta \theta}$ dominates the forward energy transfers from the largest wavelengths $\lambda_x > 10$, and the other terms show relatively small contributions, indicating forward energy transfers towards smaller wavelengths. 
The distinct differences in the behaviours of $tr^{x,(i)}_{kt}$ and $tr^{x,(i)}_{\theta \theta}$ is that the fourth term of the temperature fluctuation transport~$tr^{x,(4)}_{\theta \theta}$ plays a significant role compared to the counterpart of the kinetic energy transport~$tr^{x,(4)}_{kt}$. As shown in Fig.~\ref{fig:trxi_nw}(b), the magnitudes of the positive and negative peaks of $tr^{x,(4)}_{\theta \theta}$ are comparable to those of the third term~$tr^{x,(3)}_{\theta \theta}$, which is a clearly larger contribution than that by~$tr^{x,(4)}_{kt}$. Due to the significant peak of $tr^{x,(4)}_{\theta \theta}$ in the smallest wavelength range the top of the positive peak of the overall $tr^x_{\theta \theta}$ profile is somewhat plateau-like and the maximum location is at $\lambda_x = 0.25$ ($\lambda_x^+ \approx 60$). Therefore, the difference between the overall interscale transports $tr^x_{kt}$ and $tr^x_{\theta \theta}$ results from such significant contribution by $tr^{x,(4)}_{\theta\theta}$ in the temperature fluctuation transport.

\begin{figure}
    \centering
    \includegraphics[width=1\hsize]{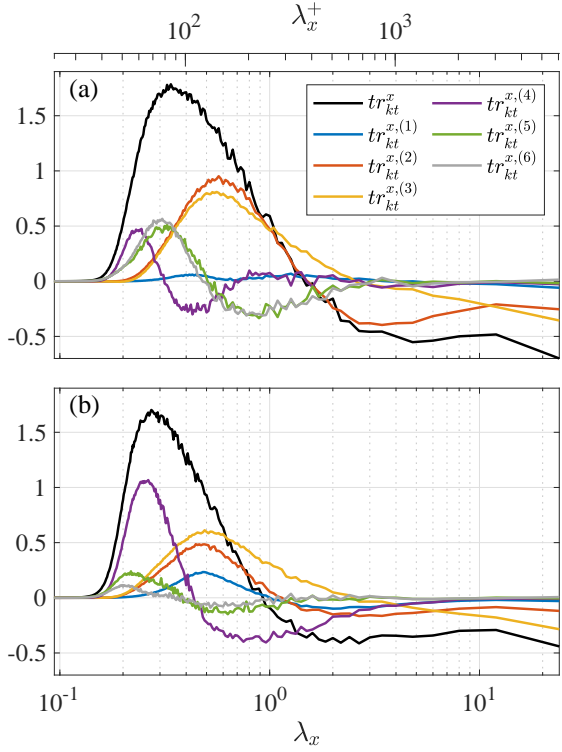}
    \caption{Comparisons of different terms in (a) $tr^x_{kt}$ and (b) $tr^x_{\theta \theta}$ at the channel centre, presented in the same manner as in Fig.~\ref{fig:trxi_nw}.}
    \label{fig:trxi_centre}
\end{figure}

At the channel centre, the difference between the behaviours of the kinetic energy transport $tr^x_{kt}$ and the temperature-fluctuation transport $tr^x_{\theta \theta}$ is not as distinct as in the near-wall region as already shown in Fig.~\ref{fig:budget_nw_cc}(b), but their ingredients present certainly different behaviours. As show in Fig.~\ref{fig:trxi_centre}(a), in the kinetic energy transport the second and third terms play the role of transferring energy from the largest wavelengths to the middle $\lambda_x$ range, the fifth and sixth terms further transfers the energy to further smaller wavelengths, and the fourth term takes the role, again, of the energy transfer in the smallest wavelength range. In the temperature-fluctuation transfer $tr^x_{\theta \theta}$, on the other hand, while the second and third terms are shown to dominate forward energy transfers in the largest $\lambda_x$ range similarly to the kinetic energy transport $tr^x_{kt}$, the contributions by the fifth and sixth terms are not significant and the fourth term is shown to take the role of transferring the energy supplied by $tr^{x,(2)}_{\theta \theta}$ and $tr^{x,(3)}_{\theta \theta}$ in the middle $\lambda_x$ range further towards smallest wavelengths, as shown in Fig.~\ref{fig:trxi_centre}(b). Such energy transfers in the small wavelength range is in contrast to the kinetic energy transport, in which the energy supply to the smallest wavelengths is accounted for by composite contributions by $tr^{x,(4)}_{kt}$, $tr^{x,(5)}_{kt}$, and $tr^{x,(6)}_{kt}$. 
It might also be worth pointing out here that each term in the kinetic energy interscale transport is decomposed into the transports of $\ave{u^2}$, $\ave{v^2}$, and $\ave{w^2}$ as
\begin{align}
    tr^{x,(i)}_{kt}=tr^{x,(i)}_{uu}+tr^{x,(i)}_{vv}+tr^{x,(i)}_{ww},
\end{align}
and while $tr^{x,(i)}_{uu}$ is dominant in all $tr^{x,(i)}_{kt}$ terms at the near-wall location $y^+=16$, the contributions by $tr^{x,(i)}_{uu}$, $tr^{x,(i)}_{vv}$, and $tr^{x,(i)}_{ww}$ are comparable at the channel centre in $tr^{x,(4)}_{kt}$, $tr^{x,(5)}_{kt}$, and $tr^{x,(6)}_{kt}$, which are responsible to the forward energy transfers at relatively small wavelengths.

\subsection{Spectral Analysis on Wall Shear Stress and Wall Heat Flux}

As observed in Fig.~\ref{fig:EUV_VT}, the integrated streamwise spectrum of the turbulent heat transfer $\mathcal{E}^x_{-v\theta}$ was shown to have its peak at relatively small $\lambda_x$ as compared to the integrated Reynolds shear stress cospectrum $\mathcal{E}^x_{-uv}$, indicating that the streamwise length scales at which turbulent heat transfer mainly occurs are relatively smaller than those of turbulent momentum transfer. Now, we address this point by investigating the transport equations of these integrated spectra $\mathcal{E}^x_{-uv}$ and $\mathcal{E}^x_{-v\theta}$, similarly to the analysis in the previous section. Integrating the transport equations of $E^x_{-uv}$ across the channel one obtains the transport equations of $\mathcal{E}^x_{-uv}$ as
\begin{subequations} \label{eq:EUVte}
\begin{align}
    \frac{{\rm D} \mathcal{E}^x_{-uv}}{{\rm D}t} &= 
    \mathcal{P}^x_{-uv} - \varXi^x_{-uv} 
    + \mathcal{W}^x_{-uv} + \mathcal{T}^x_{-uv},
    % \frac{D \mathcal{E}^x_{-v\theta}}{Dt} &= 
    % \mathcal{P}^x_{-v\theta} - \varXi^x_{-v\theta} 
    % + \mathcal{W}^x_{-v\theta} + \mathcal{T}^x_{-v\theta},
\end{align}
where $\mathcal{P}^x_{-uv}$, $\varXi^x_{-uv}$, $\mathcal{W}^x_{-uv}$, and $\mathcal{T}^x_{-uv}$ are the production, viscous dissipation, pressure work, and the interscale transport terms defined as
\begin{align}
    \mathcal{P}^x_{-uv} = \int^1_0 pr^x_{-uv} \mathrm{d}y, &\quad
    \varXi^x_{-uv} = \int^1_0 \xi^x_{-uv} \mathrm{d}y, \\
    \mathcal{W}^x_{-uv} = \int^1_0 \phi^x_{-uv} \mathrm{d}y, &\quad \mathcal{T}^x_{-uv} = \int^1_0 tr^x_{-uv} \mathrm{d}y.
\end{align}
\end{subequations}
Note here that the terms corresponding to the viscous diffusion $d^{\nu,x}_{-uv}$ and the turbulent transport $d^{t,x}_{-uv}$ do not appear in the above equations as their integrations across the channel are zero at any wavelength. The transport equation of $\mathcal{E}^x_{-v\theta}$ is also obtained by integrating the $E^x_{-v\theta}$ transport equation as 
\begin{align} \label{eq:EVTte}
    \frac{{\rm D} \mathcal{E}^x_{-v\theta}}{{\rm D}T} = 
    \mathcal{P}^x_{-v\theta} - \varXi^x_{-v\theta} 
    + \mathcal{W}^x_{-v\theta} + \mathcal{T}^x_{-v\theta},
\end{align}
and the terms on the right-hand side are defined as the integration of the corresponding terms in Eq.~(\ref{eq:eitte}) similarly in Eq.~(\ref{eq:EUVte}b,c). 

\begin{figure}
    \centering
    \includegraphics[width=1\hsize]{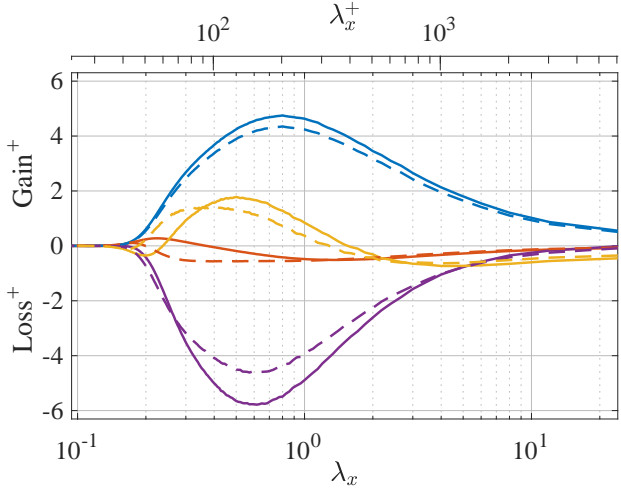}
    \caption{Comparison between the transport budgets of $\mathcal{E}^x_{-uv}$ and $\mathcal{E}^x_{-v\theta}$: (blue) productions, (red) viscous dissipations, (yellow) interscale transports, and (purple) pressure works; solid lines, $\mathcal{E}^x_{-uv}$ transport; dashed lines, $\mathcal{E}^x_{-v\theta}$ transport. The values of the terms in the $\mathcal{E}^x_{-uv}$ and $\mathcal{E}^x_{-v\theta}$ transport equations are scaled by $u_\tau^{\ast 3}$ and $u_\tau^{\ast 2} T_\tau^\ast$, respectively.}
    \label{fig:BUDGET}
\end{figure}

The budgets of the transport equations of $\mathcal{E}^x_{-uv}$ and $\mathcal{E}^x_{-v\theta}$ are compared in Fig.~\ref{fig:BUDGET}. As shown here, the production spectra $\mathcal{P}^x_{-uv}$ and $\mathcal{P}^x_{-v\theta}$ show similar profiles with their peak locations at same wavelength $\lambda_x =0.8$ (corresponding to $\lambda_x^+ \approx 200$). It should be noted here that the production spectra $pr^x_{-uv}$ and $pr^x_{-v\theta}$ both depend on the wall-normal turbulent energy spectrum $E^x_{vv}$:
\begin{align}
    pr^x_{-uv} = E^x_{vv} \frac{\mathrm{d}U}{\mathrm{d}y}, \\%\quad\quad
    pr^x_{-v\theta} = E^x_{vv} \frac{\mathrm{d}\varTheta}{\mathrm{d}y}.
\end{align}
Hence, the scales at which the productions of $\mathcal{E}^x_{-uv}$ and $\mathcal{E}^x_{-v\theta}$ mainly take place are both determined by the $E^x_{vv}$ distribution. 

As shown in Fig.~\ref{fig:BUDGET}, the budgets of the transport equations of $\mathcal{E}^x_{-uv}$ and $\mathcal{E}^x_{-v\theta}$ are basically similar. The productions are mainly balanced by the pressure work spectra $\mathcal{W}_{-uv}$ and $\mathcal{W}_{-v\theta}$, and the profiles of $\mathcal{W}^x_{-uv}$ and $\mathcal{W}^x_{-v\theta}$ show their peak magnitude at the same wavelength $\lambda_x \approx 0.6$ ($\lambda_x^+ \approx 150$) similarly to those of the productions. The viscous dissipation spectra $\varXi^x_{-uv}$ and $\varXi^x_{-v\theta}$ present only minor contributions. 

The difference between the budgets of the $\mathcal{E}^x_{-uv}$ and $\mathcal{E}^x_{-v\theta}$ equations are, again, observed in the behaviours of the interscale transport terms $\mathcal{T}^x_{-uv}$ and $\mathcal{T}^x_{-v\theta}$. As can be seen in Fig.~\ref{fig:BUDGET}, both $\mathcal{T}^x_{-uv}$ and $\mathcal{T}^x_{-v\theta}$ transfer energies from relatively large wavelengths $\lambda_x>2$ ($\lambda_x^+>500$) to smaller wavelengths around $\lambda_x \approx 0.3$--0.5, but $\mathcal{T}^x_{-v\theta}$ is shown to transfer energy to smaller wavelengths than $\mathcal{T}^x_{-uv}$. Such behaviours of the interscale transports are, therefore, attributable to the difference between the profiles of $\mathcal{E}^x_{-uv}$ and $\mathcal{E}^x_{-v\theta}$ observed in Fig.~\ref{fig:EUV_VT}(a). 

\begin{figure}
    \centering
    \includegraphics[width=1\hsize]{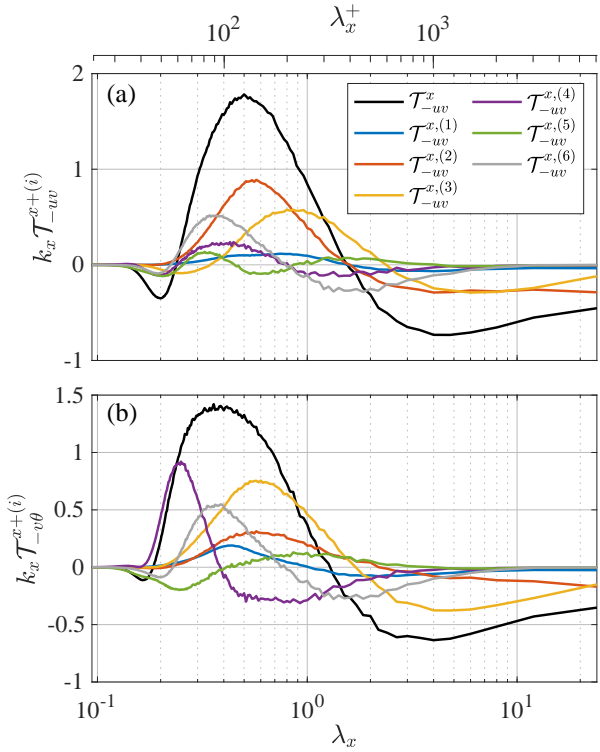}
    \caption{Comparisons between different terms in the integrated interscale transport term (a) $\mathcal{T}^x_{-uv}$ and (b) $\mathcal{T}^x_{-v\theta}$. The values are scaled by $u_\tau^{\ast 3}$ and $u_\tau^{\ast 2} T_\tau^{\ast 2}$, respectively, and the line colours in the panel~(b) represent the same terms as in the panel~(a).}
    \label{fig:TRXi_uv_vt}
\end{figure}

Similarly to Eqs.~(\ref{eq:dec_Tr}), the interscale fluxes $Tr^x_{-uv}$ and $Tr^x_{-v\theta}$ are decomposed into six different terms representing the effects of the large- or small-scale velocity gradients in different directions as
\begin{widetext}
\begin{subequations}
\begin{align}
    Tr^x_{-uv} &=
    \underbrace{ \ave{u^s u^s \pd{v^L}{x}}
                +\ave{v^s u^s \pd{u^L}{x}}}_{Tr^{x,(1)}_{-uv}} \:\:\:
    \underbrace{+\ave{u^s v^s \pd{v^L}{y}}
                +\ave{v^s v^s \pd{u^L}{y}}}_{Tr^{x,(2)}_{-uv}} \:\:\:
    \underbrace{+\ave{u^s w^s \pd{v^L}{z}}
                +\ave{v^s w^s \pd{u^L}{z}}}_{Tr^{x,(3)}_{-uv}} \nonumber \\
   &\underbrace{-\ave{u^L u^L \pd{v^S}{x}}
                -\ave{v^L u^L \pd{u^S}{x}}}_{Tr^{x,(4)}_{-uv}} \:\:\:
    \underbrace{-\ave{u^L v^L \pd{v^S}{y}}
                -\ave{v^L v^L \pd{u^S}{y}}}_{Tr^{x,(5)}_{-uv}} \:\:\:
    \underbrace{-\ave{u^L w^L \pd{v^S}{z}}
                -\ave{v^L w^L \pd{u^S}{z}}}_{Tr^{x,(6)}_{-uv}},\label{eq:dec_truv} \\
% \end{align}
% \begin{align}
    Tr^x_{-v\theta} &=
    \underbrace{ \ave{\theta^s u^s \pd{v^L}{x}}
                +\ave{v^s u^s \pd{\theta^L}{x}}}_{Tr^{x,(1)}_{-v\theta}} \:\:\:
    \underbrace{+\ave{\theta^s v^s \pd{v^L}{y}}
                +\ave{v^s v^s \pd{\theta^L}{y}}}_{Tr^{x,(2)}_{-v\theta}} \:\:\:
    \underbrace{+\ave{\theta^s w^s \pd{v^L}{z}}
                +\ave{v^s w^s \pd{\theta^L}{z}}}_{Tr^{x,(3)}_{-v\theta}} \:\:\: \nonumber \\
   &\underbrace{-\ave{\theta^L u^L \pd{v^S}{x}}
                -\ave{v^L u^L \pd{\theta^S}{x}}}_{Tr^{x,(4)}_{-v\theta}} \:\:\:
    \underbrace{-\ave{\theta^L v^L \pd{v^S}{y}}
                -\ave{v^L v^L \pd{\theta^S}{y}}}_{Tr^{x,(5)}_{-v\theta}} \:\:\:
    \underbrace{-\ave{\theta^L w^L \pd{v^S}{z}}
                -\ave{v^L w^L \pd{\theta^S}{z}}}_{Tr^{x,(6)}_{-v\theta}}.\label{eq:dec_trvt}
\end{align}
\end{subequations}
\end{widetext}
Hence, the integrated interscale transport terms $\mathcal{T}^x_{-uv}$ and $\mathcal{T}^x_{-v\theta}$ are also decomposed into the six different corresponding terms $\mathcal{T}^{x,(i)}_{-uv}$ and $\mathcal{T}^{x,(i)}_{-v\theta}$ defined as
\begin{subequations}
\begin{align}
    \mathcal{T}^{x,(i)}_{-uv} = \int^1_0 tr^{x,(i)}_{-uv} \mathrm{d}y, \quad 
    \text{where} \quad tr^{x,(i)}_{-uv}=-\pd{Tr^{x,(i)}_{-uv}}{k_x}, \\
    \mathcal{T}^{x,(i)}_{-v\theta} = \int^1_0 tr^{x,(i)}_{-v\theta} \mathrm{d}y, \quad 
    \text{where} \quad tr^{x,(i)}_{-v\theta}=-\pd{Tr^{x,(i)}_{-v\theta}}{k_x}.
\end{align}
\end{subequations}
Their contributions are compared in Fig.~\ref{fig:TRXi_uv_vt}. As one can see in the panel~(a), the terms playing major roles in the Reynolds shear stress transport are the second, third, and sixth terms, which all present forward interscale energy transfers in different wavelength ranges: the third term $\mathcal{T}^{x,(3)}_{-uv}$ accounts for the energy transfers in the largest wavelength range ($\lambda \approx 0.8$--0.9), similarly to the turbulent energy transport, and $\mathcal{T}^{x,(2)}_{-uv}$ and $\mathcal{T}^{x,(6)}_{-uv}$ represent those towards the middle and small ranges ($\lambda \approx 0.5$--0.6 and 0.3--0.4), respectively. 

As for the turbulent heat flux transfer $\mathcal{T}^x_{-v\theta}$, the third and the sixth terms, $\mathcal{T}^{x,(3)}_{-v\theta}$ and $\mathcal{T}^{x,(6)}_{-v\theta}$, present major contributions similarly to their counterparts in the Reynolds shear stress transport, but the differences between $\mathcal{T}^x_{-uv}$ and $\mathcal{T}^x_{-v\theta}$ are indicated by, again, the fourth term $\mathcal{T}^{x,(4)}_{-v\theta}$. As can be seen in Fig.~\ref{fig:TRXi_uv_vt}, the fourth term in the interscale transport of the Reynolds shear stress $\mathcal{T}^{x,(4)}_{-uv}$ shows only minor contribution but the counterpart in interscale transport of the turbulent heat transfer $\mathcal{T}^{x,(4)}_{-v\theta}$ dominates the transports in the smallest $\lambda_x$ range, similarly to the corresponding term in the temperature-fluctuation interscale transport $tr^x_{\theta \theta}$. 

Figure~\ref{fig:TRXi_uv_vt} also shows that the second terms $\mathcal{T}^{x,(2)}_{-uv}$ and $\mathcal{T}^{x,(2)}_{-v\theta}$ behave differently. While $\mathcal{T}^{x,(2)}_{-uv}$ gives the most significant energy transfer among the ingredients of the Reynolds shear stress transport $\mathcal{T}^x_{-uv}$, the contribution by $\mathcal{T}^{x,(2)}_{-v\theta}$ in the interscale energy transport is relatively small compared to other major terms, such as $\mathcal{T}^{x,(3)}_{-v\theta}$ and $\mathcal{T}^{x,(6)}_{-v\theta}$.  Due to such differences in the contributions by the second and fourth terms, the profile of  the overall interscale transfer $\mathcal{T}^x_{-v\theta}$ presents the peak at somewhat smaller wavelengths than the Reynolds shear stress transport $\mathcal{T}^x_{-uv}$.

Through the above discussions, the interscale transport of the turbulent heat flux $\mathcal{T}^x_{-v\theta}$ has been shown to transfer more spectral energy to smaller streamwise wavelengths than the interscale Reynolds shear stress transport $\mathcal{T}^x_{-uv}$, and the fourth term has been found to play a particularly important role in the smallest $\lambda_x$ range similarly to the temperature-fluctuation transport. The fourth terms of the interscale transport $\mathcal{T}^{x,(4)}_{-uv}$ and $\mathcal{T}^{x,(4)}_{-v\theta}$ are, as seen in Eqs.~(\ref{eq:dec_truv}) and (\ref{eq:dec_trvt}), comprised of two terms representing the $x$-gradients of small-scale different velocity components or temperature:
\begin{subequations}\label{eq:TRX4}
\begin{align}
Tr^{x,(4)}_{-uv} = \underbrace{-\ave{u^L u^L \pd{v^S}{x}}}_{Tr^{x,(4\text{-I})}_{-uv}}
  \; \underbrace{- \ave{v^L u^L \pd{u^S}{x}}}_{Tr^{x,(4\text{-II}))}_{-uv}}, \\
Tr^{x,(4)}_{-v\theta} = \underbrace{-\ave{\theta^L u^L \pd{v^S}{x}}}_{Tr^{x,(4\text{-I}))}_{-v\theta}} 
 \;\underbrace{- \ave{v^L u^L \pd{\theta^S}{x}}}_{Tr^{x,(4\text{-II}))}_{-v\theta}}.  
\end{align}
\end{subequations}
\begin{figure}
    \centering
    \includegraphics[width=0.9\hsize]{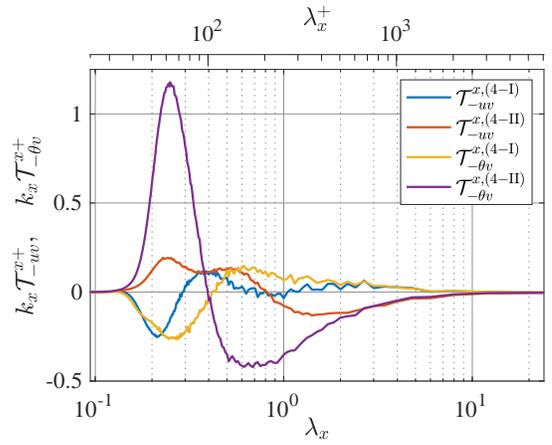}
    \caption{Comparisons between different terms in $\mathcal{T}^{x,(4)}_{-uv}$ and $\mathcal{T}^{x,(4)}_{-v\theta}$. The values are scaled by $u_\tau^{\tau 3}$ and $u_\tau^{\tau 2} T_\tau^\ast$.}
    \label{fig:TRXi4}
\end{figure}
All four terms are compared in Fig.~\ref{fig:TRXi4}. The most significant term is found to be $\mathcal{T}^{x,(4\text{-II})}_{-v\theta}$, which is related to the small-scale fluctuating temperature gradient $\partial \theta^S/\partial x$. This term is clearly responsible to the notable contribution by the fourth term $\mathcal{T}^{x,(4)}_{-v\theta}$ at the small wavelengths. This trend is similar to the interscale transport of the temperature fluctuation $tr^x_{\theta \theta}$, in that the significant energy transport to the smallest $\lambda_x$ range by $tr^x_{\theta \theta}$ is also mainly due to the term associated with $\partial \theta^S/\partial x$, as shown in the previous section. 

Thus, through this section, it has been revealed that the interscale energy transfers of temperature fluctuation at smallest $\lambda_x$ range is dominated by the term related with the streamwise temperature gradient, and the energy transport by this term has been found notably significant as compared to the corresponding term of the turbulent kinetic energy transport. Such significant interscale transport in the temperature-fluctuation spectrum transport is mainly responsible to the temperature fluctuation at smaller $\lambda_x$ than the velocity fluctuation and the certain scale-by-scale dissimilarity between the turbulent momentum and heat transfers. 

\section{Conclusion}

In the present study, a turbulent plane Couette flow with passive-scalar heat transfer has been simulated in order to investigate the interscale and spatial transports of the temperature-related statistics in comparisons with the corresponding transports of the turbulent kinetic energy and the Reynolds shear stress. Due to the similar boundary conditions in the velocity and temperature fields, the profiles of the mean streamwise velocity and the mean temperature are similar to each other, and the productions of the turbulent energy and the temperature fluctuation, therefore, also have similar profiles. In such a configuration, the distributions of the spectra and their transport budgets have been investigated in details.  

It has been found that the temperature fluctuation occurs at somewhat smaller streamwise length scales than the velocity fluctuations, although though the Prandtl number is smaller than 1 and therefore the effect of molecular diffusion is more significant in the temperature field. The turbulent heat flux spectrum is also found to be distributed in smaller streamwise wavelength range than the Reynolds shear stress cospectrum, indicating a certain scale-by-scale dissimilarity between the turbulent momentum and heat transfers. The transport budget analysis of the temperature-related spectra reveals that the interscale energy transport of the temperature fluctuation is notably significant compared to that of the turbulent kinetic energy at smallest streamwise wavelengths. Among various terms comprising the interscale energy transport, the term representing the effect by the streamwise temperature gradient at small scales is found to dominate the transport of the temperature fluctuation to the smallest wavelength range, while the counterpart in the turbulent kinetic energy transport plays only a minor role in the energy transport budget. Such difference between the interscale transports of the turbulent kinetic energy and the temperature fluctuation is attributable to the different distributions of their spectra at small scales. The same tendency is also found in the interscale transports of the Reynolds shear stress and the turbulent heat flux, which results in the spectral dissimilarity between the turbulent heat and momentum transfers.  

\begin{acknowledgments}
This work was supported by the Japan Society for the Promotion of Science (JSPS) through JSPS KAKENHI, Grant No.~JP20K14654. The numerical simulations in the present study were performed by SX-ACE supercomputers at the Cybermedia Center of Osaka University.
\end{acknowledgments}

\section*{Author Declaration}
\subsection*{Conflict of Interest}
The authors have no conflicts to disclose.

\section*{Data Availability}
The data that support the findings of this study are available from the corresponding author upon reasonable request.

% \nocite{*}
\bibliography{library_kawata}% Produces the bibliography via BibTeX.

\end{document}